\documentclass[aip,author-year,reprint]{revtex4-1}
\usepackage{amsmath,amsthm,amsfonts,amssymb}
\usepackage{graphicx}
\usepackage{subfigure}
\usepackage{longtable}
\usepackage{color}

\renewcommand{\vec}[1]{{\mathbf #1}}        
\newcommand{\ten}[1]{{\mathbf{#1}}}
\newcommand{\half}{\textstyle{\frac{1}{2}}}
\newcommand{\Id}{\ten{I}}

\begin{document}
\title{Loss of solutions in shear banding fluids in shear banding fluids driven by
  second normal stress differences}
\author{S.~Skorski}
\affiliation{ School of Physics \& Astronomy, University of Leeds,
  Leeds, LS2 9JT, U.K.} 
\author{P.~D.~Olmsted}
\email{p.d.olmsted@leeds.ac.uk}
\affiliation{ School of Physics \& Astronomy, University of Leeds,
  Leeds, LS2 9JT, U.K.}
\date{\today}
\begin{abstract}
  Edge fracture occurs frequently in non-Newtonian fluids. A similar
  instability has often been reported at the free surface of fluids
  undergoing shear banding, and leads to expulsion of the sample.  In
  this paper the distortion of the free surface of such a shear
  banding fluid is calculated by balancing the surface tension against
  the second normal stresses induced in the two shear bands, and
  simultaneously requiring a continuous and smooth meniscus.  We show
  that wormlike micelles typically retain meniscus integrity when
  shear banding, but in some cases can lose integrity for a range of
  average applied shear rates during which one expects shear
  banding. This meniscus fracture would lead to ejection of the sample
  as the shear banding region is swept through. We further show that
  entangled polymer solutions are expected to display a propensity for
  fracture, because of their much larger second normal stresses. These
  calculations are consistent with available data in the
  literature. We also estimate the meniscus distortion of a three band
  configuration, as has been observed in some wormlike micellar
  solutions in a cone and plate geometry.
\end{abstract}
\preprint{11111}
\maketitle
\section{Introduction}
Many complex fluids are dramatically influenced by shear flow, which
can easily disrupt the slow microstructural relaxation of these
fluids. Examples include polymer solutions \citep{boukany2009shear},
wormlike micellar \citep{schmitt94} and lamellar \citep{DRN93}
surfactant solutions, colloidal suspensions \citep{ChenCAZ94}, and
telechelic polymer networks \citep{michel2001ufa}, In many cases shear
flow can induce an apparent transition to a state with a different
microstructure and apparent viscosity, which can lead to macroscopic
bands of material that coexist, much like phase separation, in flow
\citep{olmstedbanding08,fieldingsoft07}.  In the most commonly
observed scenario, the system forms two or more `shear bands'; layers
of high and low shear rate material (of equal shear stress) that
coexist at volume fractions consistent with an imposed average shear
rate \citep{schmitt94,Britton.Callaghan99,Capp+97b}. As the average
shear rate is increased, the width of the high shear rate band
increases, while a constant stress is maintained (in an idealized
planar Couette geometry). The measured shear stress as a function of
applied average shear rate is the \textit{flow curve}, and contains a
broad stress plateau for shear rates in the banding range.

Shear banding can result when the constitutive relationship between
shear stress and shear strain rate, assuming homogeneous flow, has a
stress maximum and is thus non-monotonic \citep{spenley93} (see
Fig.~\ref{fig:Fig2}a below). Homogeneous flow is unstable for applied
shear rates in the region of the \textit{constitutive curve} with a
negative slope, and this instability can be resolved by adopting the
shear banding state. This constitutive instability is present in the
Doi-Edwards (DE) reptation model for entangled polymers
\citep{doiedwards} for sufficiently weak levels of Convected
Constraint Release (CCR) \citep{graham2003microscopic}, and was only
recently been observed in polymer solutions
\citep{tapadiawang03,hu2007cre,boukany07,boukany2009shear,adamsolmsted09a}.
Wormlike micelles have a similar constitutive instability, due to the
combination of reptation and micellar breakage
\citep{cates90,rehage91,spenley93}, and shear banding has been studied
in these systems for decades \citep{berretrev05}.

In many experiments on shear banding the free surface fractures
and the sample is ejected from the device. Although it is often
reported as a nuisance and anecdotally, it is widespread in both
worm-like micellar solutions \citep{BritCall97c,BPD97,LopGon} and
entangled polymer solutions
\citep{inn2005eef,sui2007iep}. Fracture and
ejection can occur at some point on the stress plateau, which
corresponds to a certain minimum width of the high shear rate
band. This is evident in the experiments of \citet{BPD97}, in which
they reported surface fracture on the stress plateau in a cone-and-plate
rheometer. They attributed this to a well-known secondary flow
instability in cone-and-plate rheometers, due to the balance of the
second normal stress difference with surface tension
\citep{tanner1983shear,Lars92b}:
\begin{equation}
  \label{eq:tanner}
  |N_2|>\frac{2\gamma}{3W},
\end{equation}
where $W$ is the maximum cone-plate separation and
$N_2=T_{yy}-T_{zz}$ is the second normal stress difference. The
balance of normal stresses with surface tension leads to a radius of
curvature $R\sim\gamma/N_2$. If this radius is too small, then the
interface must curve too tightly to fit inside a wide gap $W$, and fracture results.

\begin{figure*}[htb]
\includegraphics[width = 1\textwidth]{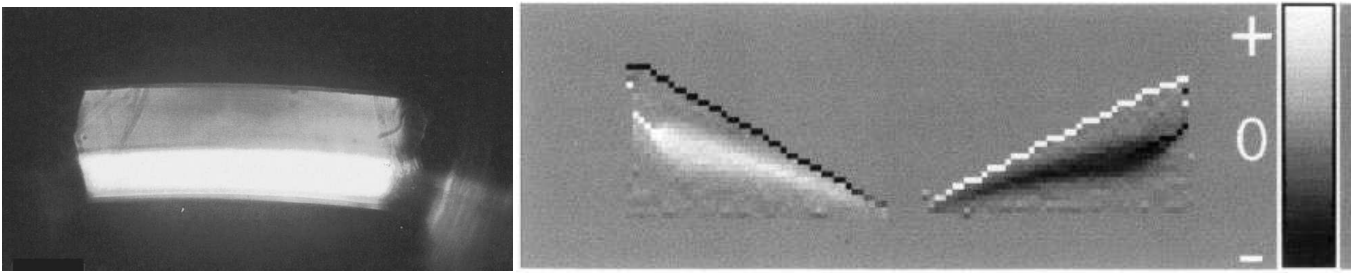}
\caption{(Left) Flow birefringence image of shear banding in a
  cylindrical Couette rheometer; the bright band is flowing at the
  higher shear rate \citep{Capp+97b}. (Right) NMR image of shear
  banding in a cone-and-plate rheometer, with a high shear rate
  central band (white or black) ( Fig. 2a \citet{BritCall97c}). A free exterior fluid surface is specified.}
\label{fig:Cells}
\end{figure*}

In this paper we propose a simple model that generalizes this idea to
incorporate shear banding, which can also address this lack of surface integrity. We
study the entire meniscus shape, as determined by second normal
stresses in the two shear bands and contact angles, and find a range
of conditions under which the meniscus cannot maintain the shape
demanded by mechanical equilibrium. As the shear bands change size,
for increasing imposed average shear rate, one or the other band
develops a width that cannot support the meniscus curvatures demanded
by the normal stress balances in both shear bands and continuity.  We
establish the conditions for mechanically stable bands, and illustrate
this behaviour using non-monotonic forms of the 
\citet{johnson77} and  \citet{Gies82} constitutive models.

We compare this explicitly to data in the literature. There is
significant data on wormlike micelles, which are for the most part
consistent with our calculations.  We also compare with recent work on
entangled polymer solutions
\citep{inn2005eef,sui2007iep,Tapadia2004Nonlinear-flow-}, which are
now known to shear band and have generated much discussion in the
literature. We show that the larger apparent viscosities of entangled
polymer solutions leads to more unstable shear bands; we hope that
this helps resolve some of the contradictory results in the literature
as to whether or not shear banding is intrinsic to the constitutive
behavior or due to a meniscus distortion. In our view shear banding
can initiate the distortion and fracture seen in some recent experiments
\citep{inn2005eef,sui2007iep}.

In Section \ref{sec:model-stab-cond} we present our model, flow
geometry, and mechanical balance conditions. In Section
\ref{sec:menisc-shape-stab} we calculate the meniscus shapes, and
derive meniscus integrity criteria that depend only on the second normal
stresses, the gap size, the surface tension, and the contact angle. In
Section \ref{sec:application-model} we compare our calculations with
experiments on wormlike micelles and polymer solutions, and we
conclude in Section \ref{sec:conclusion}. Appendix
\ref{sec:const-equat} collects the relevant information for the
Giesekus and Johnson-Segalman models, and Appendix
\ref{sec:three-band-conf} contains the details for calculations in
which a center high (or low) shear rate band is sandwiched between two
low (or high) shear rate bands.
\section{Model and Meniscus Integrity Conditions}\label{sec:model-stab-cond}
\textit{Constitutive Equations---}We consider an incompressible fluid
obeying the following relationship between shear stress and shear
strain rate
\begin{equation}
\ten{T} = -p \Id + 2 \eta \ten{D} + \ten{\Sigma}\;,
\label{eqn:totalstress}
\end{equation}
where $\Id$ is the identity tensor, $\ten{D} \equiv \half\left[\nabla
  \vec{v} + (\nabla \vec{v})^T\right]$, $p$ is the isotropic pressure
determined by incompressibility ($\nabla \cdot \vec{v} = 0$), $\eta$
is an assumed Newtonian viscosity (due to solvent or other fast modes),
and $\vec{v}$ is the velocity field. The stress tensor $\ten{\Sigma}$
is an additional viscoelastic stress that has its own dynamical
equation of motion. We will illustrate our results using the
Johnson-Segalman (JS) and Giesekus models, whose details are presented in
the Appendix.  We will consider steady creeping flow, so
\begin{equation}
\nabla \cdot \ten{T} = 0\;.
\label{eqn:delT}
\end{equation}

Shear banding usually develops only two bands, although occasionally
more complex structures are seen, such as a three band configuration
in cone-and-plate (Fig.~\ref{fig:Cells}). The stress gradient in
cylindrical Couette flow typically ensures that two bands develop with
the high shear rate phase near the inner cylinder \citep{olmsted99a},
while the much weaker stress gradient of cone-and-plate flow could
explain the more complex structures \citep{BCAdams08}. Alternatively,
\citet{KumarL00} showed  that unidirectional shearing flow, in 
the cone and plate geometry, of adjacent fluids with different normal stresses is incompatible 
with momentum balance. Based on this three band state, we calculate the meniscus distortion of two band and
three band configurations. We specialize to a planar Couette geometry
for all calculations. We may sometimes refer to a particularly shear banding configuration as `unstable' if the mechanical balance condition does
not yield a physical reasonable interface. However, we emphasize throughout that we do not calculate the conditions for dynamic 
instability, but rather the conditions under which a meniscus solution is consistent with 
momentum balance. This `instability' or lack of solution may may, of course, be preempted by an instability due to secondary flow arising from purely dynamical considerations.

\begin{figure*}[htb]
\includegraphics[width = 1\textwidth]{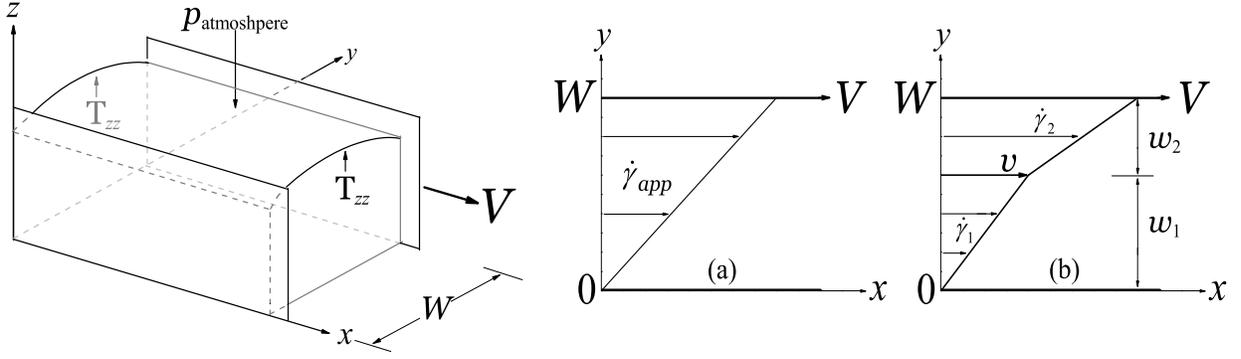}
\caption{(Left) Schematic diagram of planar Couette flow. The plate in the $y=0$ plane is stationary whilst a parallel plate a distance $W$ away moves with velocity $V$. The free surface at large $z$ is open to the atmosphere. (Right) Velocity profiles for an applied average shear rate
  $\dot{\gamma}_{\textrm{app}}=\frac{V}{W}$:  (a) unbanded or (b)
  shear banded. For banded flow $\dot{\gamma}_1=\frac{v}{w_1}$ and
  $\dot{\gamma}_2=\frac{V-v}{w_2}$, where $w_1$ and $w_2$ are the
  widths of the two shear bands.}
\label{fig:Fig1}
\end{figure*}

\textit{Shear Flow and Mechanical Balance Conditions--} We consider steady
laminar flow between two parallel plates a distance $W$ apart; one plate is stationary and the other plate moves with a constant
speed $V$ in the $x$ direction (see Fig.~\ref{fig:Fig1}). The flow
gradient and shear rate only vary in the $y$ direction, $\partial
v_x/\partial y = \dot{\gamma}(y)$. The applied average shear rate is
$\dot{\gamma}_{\textrm{app}}=V/W$. The free surface (meniscus), specified by a height $h(y)$, is in
the z direction, and we assume no-slip boundary conditions at the
solid walls.

If the free surface is considered at all it is usual for the meniscus to be flat (parallel to the plane $z=0$). Curvature of the free surface is problematic; stress balance at the surface implies that the base flow near the surface will, generally, no longer be given by $\vec{v}=\big(\dot{\gamma}y,0,0\big)$. For $h^\prime\!\ll\!1$ this may not be a problem; otherwise secondary flows will develop and, generally, we expect hydrodynamic instabilities to preempt our estimates of non-existence of surface integrity. This is discussed in the Appendix.
  
\begin{figure*}[htb]
\includegraphics[width = 1\textwidth]{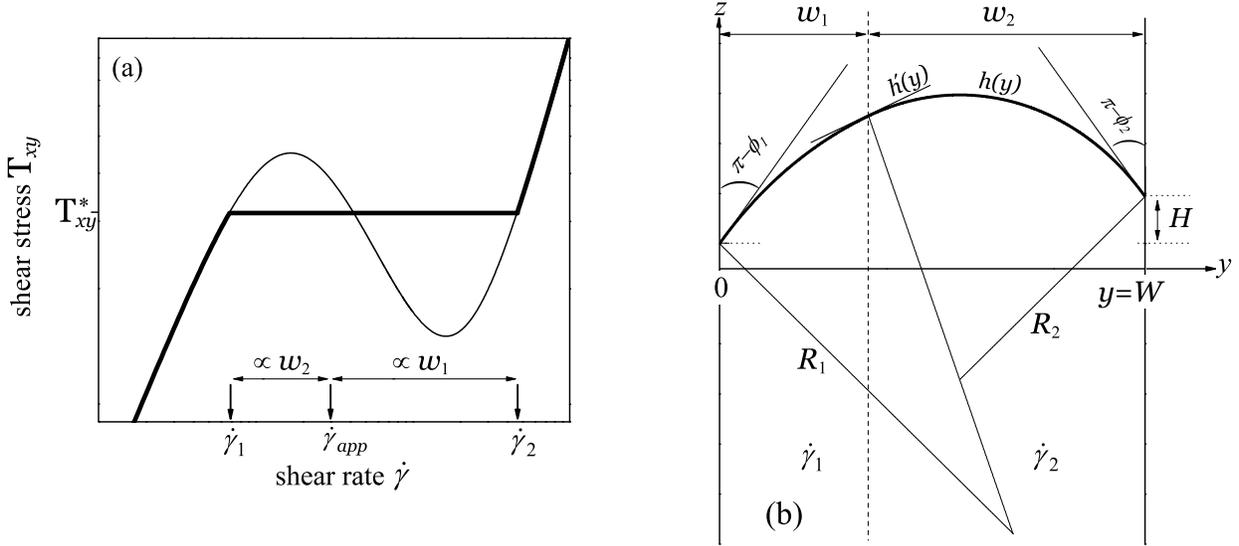}
\caption{(a) Flow curve (thick line) and constitutive curve (thin
  line) for shear banding flows. Banding occurs on the stress plateau
  at ${\rm T}^{\ast}_{xy}$, for applied shear rates
  $\dot{\gamma}_{\textrm{app}}$ such that
  $\dot{\gamma}_1\leq\dot{\gamma}_{\textrm{app}}\leq\dot{\gamma}_2$. A
  `lever' rule relates the widths of the shear bands to the applied
  shear rate, $w_1 = W
  (\dot{\gamma}_2-\dot{\gamma}_{\textrm{app}})/(\dot{\gamma}_2-\dot{\gamma}_1)$
  and $w_2 = W-w_1$.
  (b) Profile of the fluid surface (meniscus) between flat plates
  (flow in the $\hat{x}$ direction). Each band has a circular profile,
  and the surface is continuous and differentiable; $\phi_1$ and $\phi_2$ are the contact
  angles at the two plates. The contact lines between fluid and plates move up and down as the widths of the shear bands alter; $H$ is the height difference between the contact lines. Note that the curvatures may be positive or negative.}
\label{fig:Fig2}
\end{figure*}

The fluid is assumed to have a non-monotonic constitutive relation,
with shear bands that form at shear rates $\dot{\gamma}_1$ and
$\dot{\gamma}_2$ two shear bands Fig.~\ref{fig:Fig2}(a).  The shear
band widths $w_1$ and $w_2$ are given by
$\hat{w}_1=-(\dot{\gamma}_{\textrm{app}} -
\dot{\gamma}_2)/(\dot{\gamma}_2-\dot{\gamma}_1)$ and
$\hat{w}_2=(\dot{\gamma}_{\textrm{app}}-\dot{\gamma}_1)/(\dot{\gamma}_2-\dot{\gamma}_1)$,
where here and throughout this paper all lengths with a carat
$\hat{\phantom{.}}$ been normalized relative to the plate gap size
$W$.  The shear bands are assumed to partition, and vary in thickness,
along the flow gradient direction.

Stress balance (Eq.~\ref{eqn:delT}) implies that $\text{T}_{xy}$ and
$\text{T}_{yy}$ are the same in each shear band. This implies that
$\Sigma_{xy}^{(1)}=\Sigma_{xy}^{(2)}$, where $^{(1)}$ and $^{(2)}$ refer
to the two shear bands. Moreover, 
\begin{equation}
  \label{eq:Sigmayy}
  \Sigma_{yy}^{(1)}-p_1 = \Sigma_{yy}^{(2)}-p_2.
\end{equation}
Since the normal stresses will generally be different in the two
bands, the pressures $p_1$ and $p_2$ will differ. These pressures must then balance,
together with  $\Sigma_{zz}$, against the curvature of the meniscus
and the atmospheric pressure $p_{atm}$:
\begin{align}
\text{T}_{zz}^{(i)}=-p_i+{\Sigma}_{zz}^{(i)}(\gamma_i) &=-p_{atm}-\frac{\gamma_s}{R_i}
\quad (i=1,2),
\label{eqn:LaPlace}
\end{align}
where $R_i$ is the radius of curvature of the meniscus in the $i$th
band and $\gamma_s$ is the surface tension. From
Eq.~(\ref{eq:Sigmayy}) the difference in $\text{T}_{zz}$ between the
two bands is given by the difference in the second normal stress
differences,
$\text{T}_{zz}^{(1)}-\text{T}_{zz}^{(2)}=N_2^{(2)}-N_2^{(1)}$. Making
use of this and Eq.~(\ref{eqn:LaPlace}), we can relate the second
normal stress differences to the radii of curvature of the two bands
\begin{equation}
\frac{\Delta N_2}{\gamma_s} = \frac{1}{R_2}-\frac{1}{R_1},
\label{eqn:deltaN}
\end{equation}
where $\Delta N_2=N_2^{(2)}-N_2^{(1)}$. This can be easily calculated
for a given constitutive relation, and together with the surface
tension defines a characteristic `elasto-capillary' length for the
shear banding configuration:
\begin{equation}
\zeta=\frac{\gamma_s}{|\Delta N_2|}.\label{eq:elastocap}
\end{equation}

The three equations (\ref{eq:Sigmayy},~\ref{eqn:LaPlace}) relate
four unknown quantities: the pressures and meniscus curvatures in
each band. By eliminating the pressures we can relate curvature radii,
but more information is needed to absolutely determine the shape. This
will follow by constructing a continuous and smooth meniscus.  The
balance at the meniscus, Eq.~\eqref{eqn:LaPlace}, determines the
pressure in each band in terms of the meniscus curvature. We will find
below that the curvatures must change in order to maintain a physical
meniscus, which thus determines the pressure in each band.

\section{Meniscus Shape and Integrity}\label{sec:menisc-shape-stab}
\subsection{General Shape (two bands)}

We ignore deviations due to complex flows near the contact line, and
approximate the meniscus of each band as the arc of a circle of radius
$ R_i$, which may be positive or negative. We demand continuity of the surface and its tangent at the
interface between two bands. The conditions above can be fulfilled either by:
\begin{enumerate}
\item[(i)]
the contact angles being fixed and the fluid adopting a height $H$ difference between the contact lines along each plate (movement of the contact lines has been observed by \citet{Crawley1977geometry}); this height difference will vary in response to the widths of the shear bands;
\item[(ii)]
the contact lines being pinned such that $H$ is fixed and the contact angles vary in response.
\end{enumerate}
In both cases the meniscus profiles are governed by the same equations.
\begin{widetext}
The height profiles of the two bands are given by
\begin{equation}
h(y)=
\begin{cases}
R_1\left(\sqrt{1-\left(\dfrac{y}{R_1}+\cos\phi_1\right)^2}-\sin\phi_1\right) & \quad (0\leq y\leq w_1)\\[3ex]
 R_2\left(\sqrt{1-\left(\dfrac{y-W}{R_2}-\cos\phi_2\right)^2}-\sin\phi_2\right)+H & \quad (w_1<y\leq W)\:,\
\end{cases}
\label{eqn:height}
\end{equation}
where $\phi_1$ and $\phi_2$ are the contact angles at either
wall, and $h(0)=0$ and $h(W)=H$.
Continuity of $h$ at $y=w_1$ requires
\begin{equation}
H=(R_1-R_2)\sqrt{1-\Big(\frac{w_1}{R_1} + \cos\phi_1\Big)^2} +
R_2\sin\phi_2-R_1\sin\phi_1\;, 
\label{eqn:topC}   
\end{equation}
and continuity of $h^\prime$ at $y=w_1$ requires
\begin{equation}
\frac{w_1}{R_1}+\frac{w_2}{R_2}=-\cos\phi_1-\cos\phi_2\;.
\label{eqn:topA}
\end{equation}
\subsection{Fixed Contact Angles}
We allow the surface tensions at the fluid, walls, and atmosphere interface to determine the contact angles and assume that the contact angles persist through the onset of banding.
Then Eq.~\eqref{eqn:deltaN} and Eq.~\eqref{eqn:topA} completely specify the
shape:
\vskip0.1truecm
\begin{equation}
\hat{R}_1=\frac{1}{-\cos\phi_1-\cos\phi_2 - \hat{w}_2 A} \quad \text{and} \quad \hat{R}_2=\frac{1}{-\cos\phi_1-\cos\phi_2 + \hat{w}_1 A}\;,
\label{eqn:topB}
\end{equation}
where the dimensionless distortion parameter
\begin{equation}
A\equiv \frac{W\Delta N_2}{\gamma_s}=-\frac{W}{\zeta}
\label{eqn:ParA}
\end{equation}
controls the shape. In the limit of high surface tension,
$|A|\simeq0$, both radii are equal and completely determined by the
contact angles.  
We have chosen negative
values for $A$ here since we expect $N_2^{(2)}<N_2^{(1)}\leq0$ in the
high shear rate band for most polymer and micellar solutions (a
similar analysis can be done for $A>0$). In Section
\ref{sec:application-model} we analyze  recent experiments and estimate
$-A\sim 0.8$ - $3$ for shear banding wormlike micelles and $-A\sim 3$ - $140$
for entangled polymer solutions.
\end{widetext}

\subsection{Integrity of Meniscus}
An interface solution exists under two conditions, one mathematical
and the other practical: 
\begin{itemize}
\item The curves should not have infinite slope or pass through each
  other.
This leads to the condition
\begin{equation} 
\Big|\hat{w}_2\cos\phi_1-\hat{w}_1\cos\phi_2-\hat{w}_1\hat{w}_2A\Big|<1\:. 
\label{eqn:StabConditions}
\end{equation}
\item The increase $H$ in height across the gap should not be
  large enough for the sample to climb out of the cell. This obviously
  depends on the loading conditions, and can generally preempt the
  mathematical condition above.
\end{itemize}
\subsection{Equal Contact Angles}
\begin{figure*}[!htb]
\includegraphics[width = 1\textwidth]{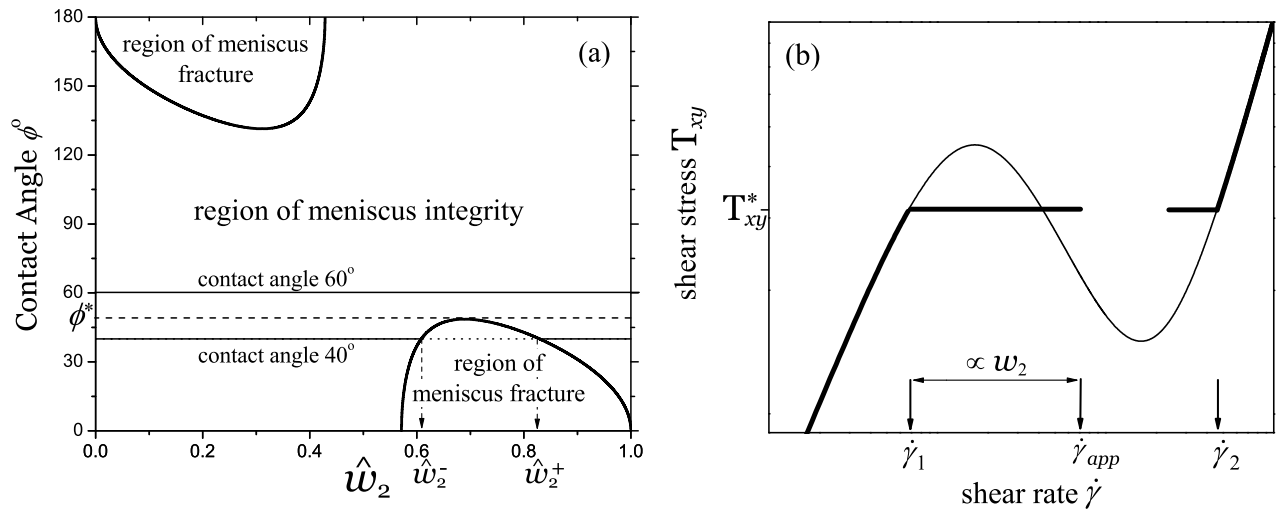}
\caption{(a) Integrity contours for equal contact angles
  $\phi_2=\phi_1 =\phi$, for $A=-3.5$. For a wall and fluid combination that fixes $\phi=60^\circ$ all shear band widths are allowed and the applied shear rate will traverse the plateau. For a wall and fluid combination that fixes $\phi=40^\circ$ not all shear rates accessible; the surface of the fluid becomes fracture for some widths $\hat{w}_2$ of the high shear rate band satisfying  $\hat{w}^-_2<\hat{w}_2<\hat{w}^+_2$. All widths allow meniscus integrity for $|\cos\phi|<\cos\phi^{\ast}=0.661$. (b) Flow curve (thick line) for a wall and fluid combination with $\phi=40^\circ$, showing the inaccessible segment of the stress plateau.}
\label{fig:Fig5}
\end{figure*}
We first consider equal contact angles $\phi_1=\phi_2=\phi$, for which
the meniscus integrity condition is
\begin{equation} 
\left|(\hat{w}_2-\hat{w}_1)\cos\phi-\hat{w}_1\hat{w}_2A\right|<1, 
\label{eqn:StabCondEqual}
\end{equation}
where the criterion
\begin{equation}
-1-\cos\phi<\frac{\hat{w}_i}{\hat{R}_i}<1-\cos\phi 
\label{eqn:topD}
\end{equation}
is required for Eq.~\eqref{eqn:topC} to yield real solutions.
Fig.~\ref{fig:Fig5}a shows regions of integrity for $A=-3.5$ and two chosen cases. For a wall and fluid combination that fixes
$\phi=60^{\circ}$ the entire shear stress plateau is accessible and
supports a integral meniscus. However, for a wall and fluid combination that fixes $\phi=40^{\circ}$ some values of $w_2$ are not allowed and the corresponding applied shear rates are
not accessible on the stress plateau. For small contact angles the
low shear rate portion of the plateau is accessible, while for high
contact angles the high shear rate portion of the plateau is
accessible.  Fig.~\ref{fig:Fig5}b shows a possible flow curve for the $\phi=40^{\circ}$ case.

\begin{figure*}[htb!]
\begin{center}
{\includegraphics[width = 1\textwidth]{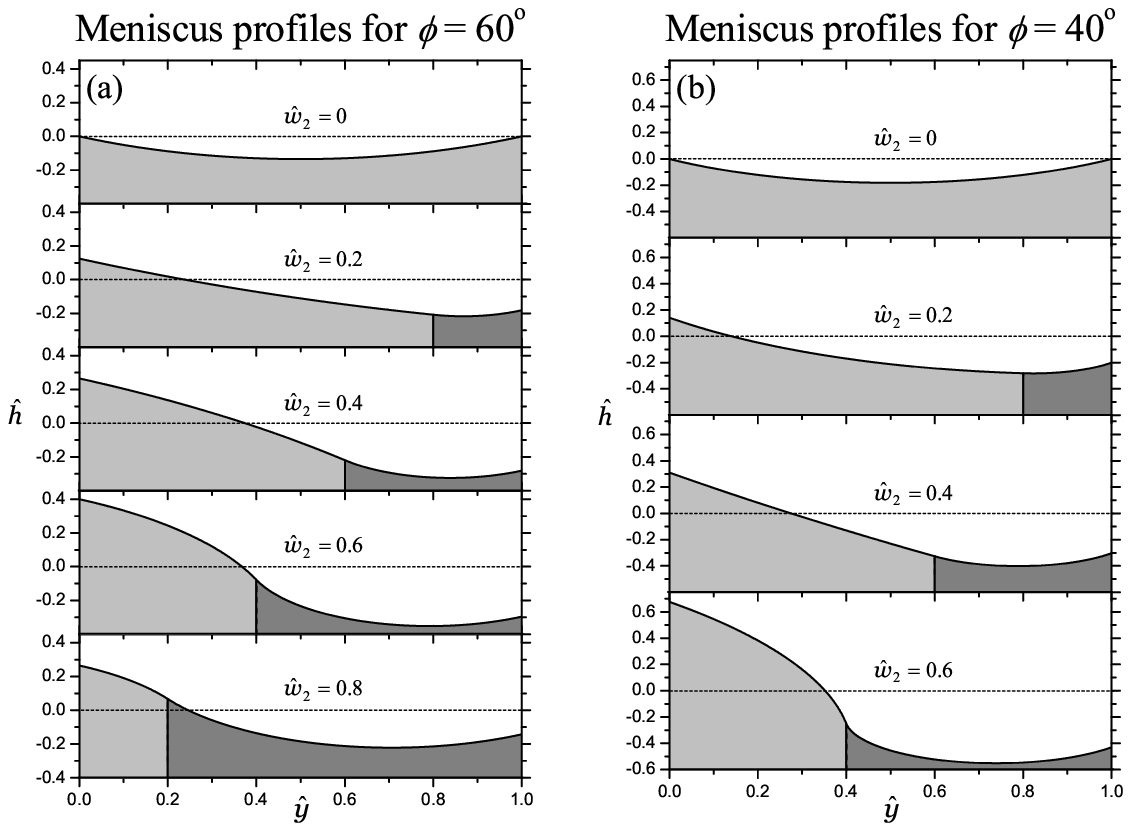}}\\
{\includegraphics[width = 1\textwidth]{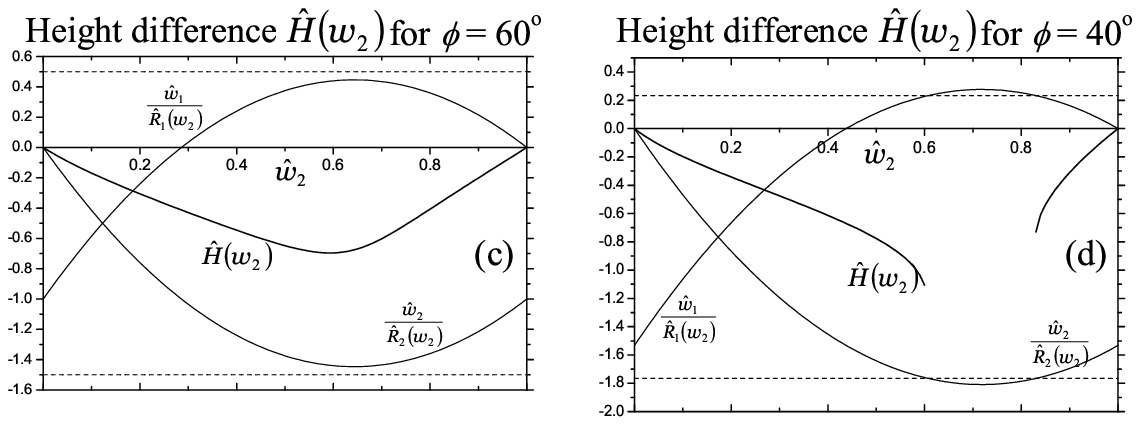}}
\end{center}
\caption{(a,c) Meniscus profiles $h(y)$; and
 (b,d) height difference $\hat{H}$
and curvatures normalized by the gap size,
  $\frac{w_i}{R_i}$,
  as a function of the high shear band width $\hat{w}_2$; for $A=-3.5$. The dotted
  lines are the integrity limits on $\frac{w_i}{R_i}$ from
  Eq.~\eqref{eqn:topD}. For the $\phi=60^\circ$ case (a,c) the surface remains integral for all shear rates in the plateau region, while for the
  $\phi=40^\circ$ case (bd) the surface of the fluid will fracture on some portion of the stress plateau.}
\label{fig:Fig4}
\end{figure*}

To understand the lack of integrity of the meniscus we must examine the
meniscus shapes. These are shown in Fig.~\ref{fig:Fig4} as a function
of increasing the applied shear rate $\dot{\gamma}_{\textrm{app}}$
across the stress plateau, for both a integral case ($\phi=60^\circ$)
and an non-integral case ($\phi=40^\circ$), for $A=-3.5$. Note that
banding first initiates when the high shear rate band develops a
non-zero width, \textit{i.e.} when  $\hat{w}_2$ grows from zero for
$\dot{\gamma}_{\textrm{app}}\geq\dot{\gamma}_1$.

Consider first the $\phi=60^\circ$ case (Fig.~\ref{fig:Fig4}a). At
$\hat{w}_2=0.2$ the contact line of the higher rate shear band has
dropped while that of the lower shear rate band has risen. Both
surfaces have negative radii of curvature. By $\hat{w}_2=0.4$ the
radius of curvature of the lower shear rate band has become
positive. The height difference between the contact lines increases and
reaches a maximum at $\hat{w}_2\simeq0.65$, but the surface still maintains
its integrity. For larger $\hat{w}_2$ the height difference decreases
until shear banding ceases at $\hat{w}_2=1$
($\dot{\gamma}_{\textrm{app}}=\dot{\gamma}_2$). For the $\phi=40^\circ$ case,
however (Fig.~\ref{fig:Fig4}b), the integrity of the surface cannot be
fulfilled when the width of the high shear rate band satisfies
$0.6<\hat{w}_2<0.83$ (see Fig.~\ref{fig:Fig4}d).As can be see in
Fig.~\ref{fig:Fig4}b, the interface develops a kink immediately
before the onset of fracture. In the region of fracture we find that
there is no continuous solution to the meniscus profiles.

Figures \ref{fig:Fig4}(cd) show the height difference $\hat{H}$, and
the surface curvatures for the different shear bands $\frac{w_i}{R_i}$
normalized by the respective band size, as a function of shear band
width $\hat{w}_2$ (or equivalently the applied shear rate
$\dot{\gamma}_{\textrm{app}}$).  The surface will maintain integrity so long as
solutions for $h'(y)$ (and consequently $H$) exist. The dotted lines
indicate between which values $\frac{\hat{w}_1}{\hat{R}_1}$ and
$\frac{\hat{w}_2}{\hat{R}_2}$ must lie in order to maintain integrity.

\begin{figure*}[htb]
\begin{center}
\includegraphics[width = 1\textwidth]{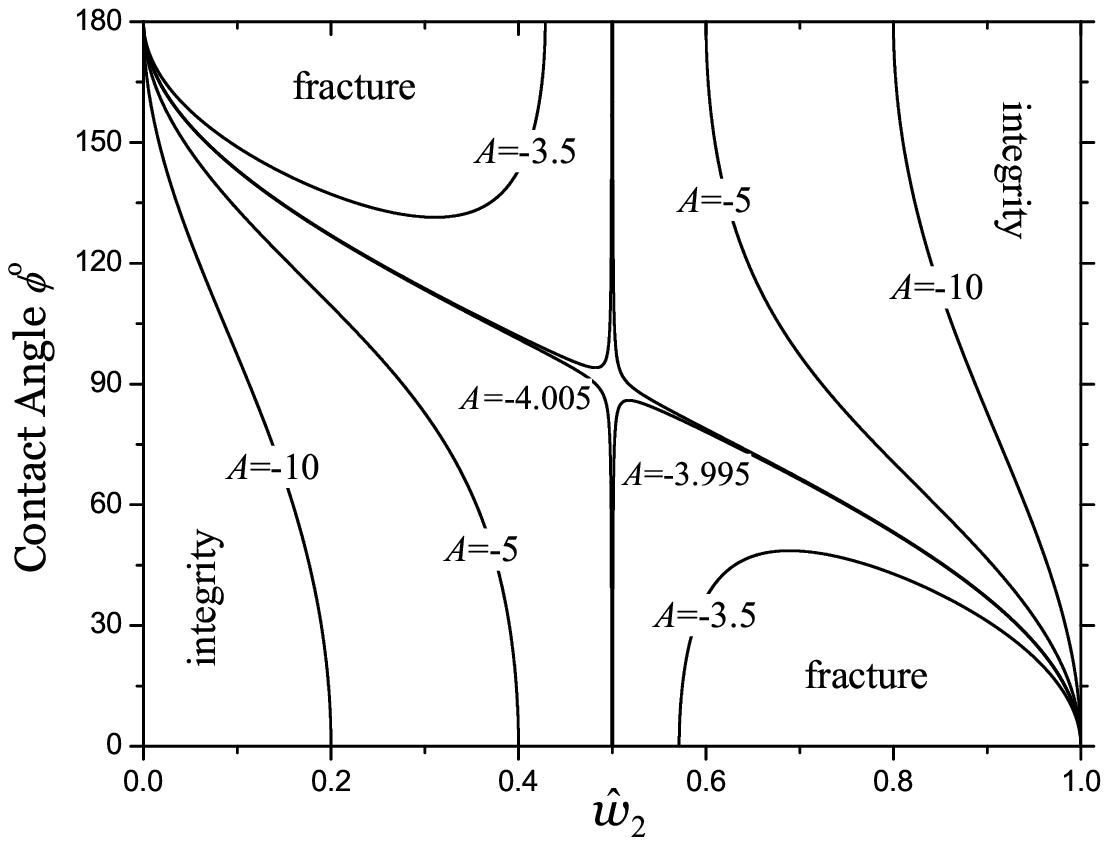}
\end{center}
\caption{Regions of integrity for equal contact
  angles $\phi$ and for different values of the distortion parameter
  $A$.}
\label{fig:Fig6}
\end{figure*}

The integrity contours for a range of distortion parameters $A$ are
shown in Fig.~\ref{fig:Fig6}, and the different regions of integrity
are summarized in Table~\ref{tab:one}. For $A>-2$ the meniscus is
always integral.
For $-4<A<-2$ the meniscus is integral for contact angles satisfying 
\begin{equation}
  \label{eq:philimit}
  \cos^2\phi\leq \cos^2\phi^{\ast}=|A|\left(1 - \frac{|A|}{4}\right).
\end{equation}
For large magnitude $|A|>4$
the meniscus looses integrity at all contact angles, for some regions of the
stress plateau. For $A<0$, as in our case, this occurs when the width
of the high low shear rate band is between the limits
\begin{equation}
  \label{eq:Wlimits}
  \hat{w}_2^{\pm} = \frac12\left(1 +
    \frac{2\cos\phi}{|A|}\right)\left[1 \pm 
      \sqrt{1 - \frac{4|A|\left(1+\cos\phi\right)}{\left(|A| +
            2\cos\phi\right)^2}}\right].
\end{equation}
These limits also apply for $-4<A<-2$, for
$|\cos\phi|\geq|\cos\phi^{\ast}|$. 
\begin{figure*}[htb!]
\begin{center}
\includegraphics[width = 1\textwidth]{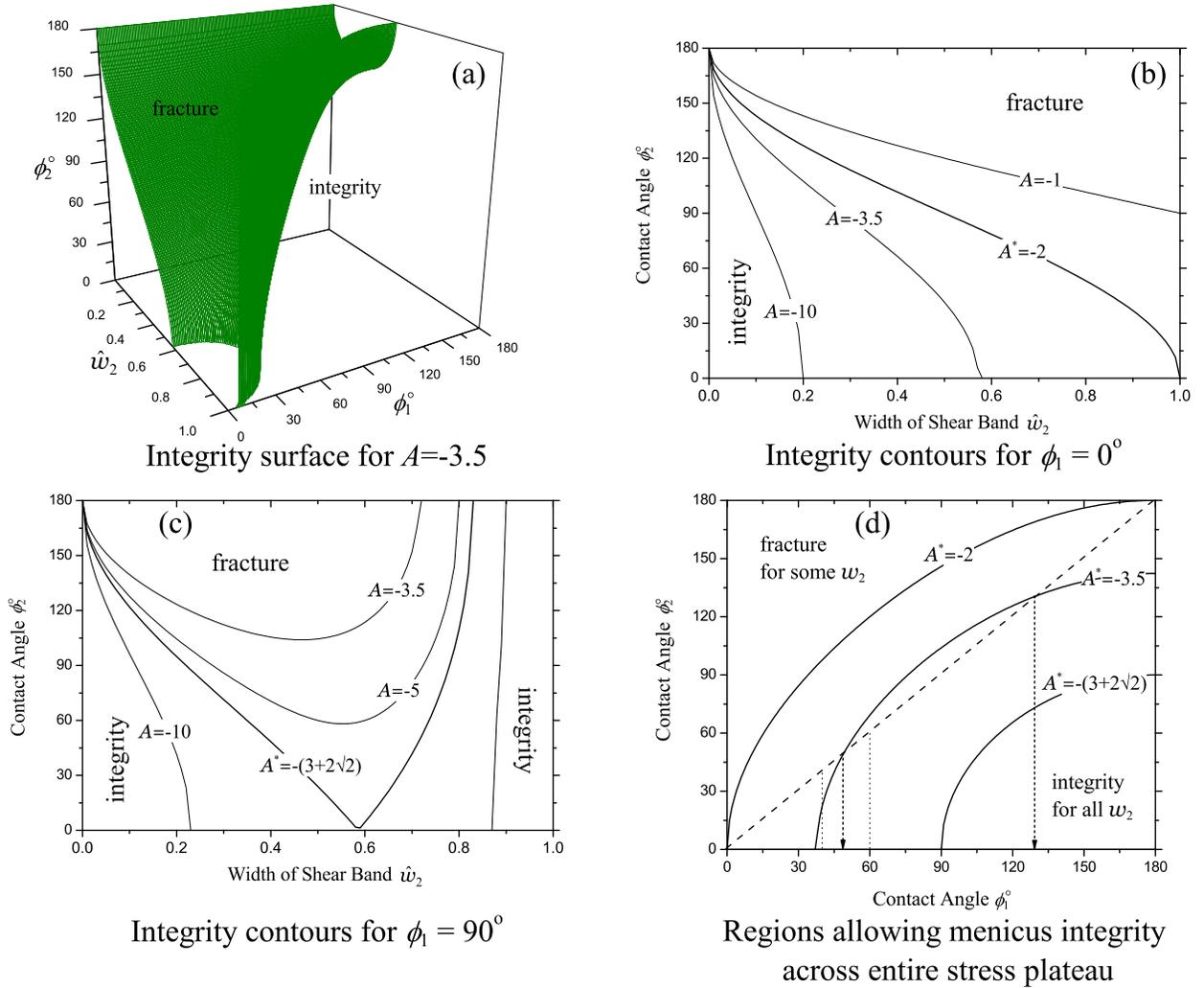}
\end{center}
\caption{(ab) Regions of integrity as a function of contact angles
  $\phi_1$ and $\phi_2$, and the width $\hat{w}_2$ of the high shear
  rate band for distortion parameter $A=-3.5$. The meniscus is integral
  for any combination of $\hat{w}_2$, $\phi_1$, and $\phi_2$ in the
  integral region. Fig.~\ref{fig:Fig5}a is the intersection of this
  integrity surface with the plane $\phi_1=\phi_2$. For $A=-1$ and
  $\phi_2<90^{\circ}$ the meniscus is integral for all  $\hat{w}_2$ (\textit{i.e.} across the entire stress
  plateau). For $A<A^*=-2$ the meniscus fractures at some shear
  band width $\hat{w}_2$ for all contact angles $\phi_2$. (c)
  Integrity contours for $\phi_2=90^{\circ}$.  The meniscus fractures at some $\hat{w}_2$ for all $\phi_2$ if $A<-(3+2\sqrt{2})$. (d) Contours
  specified by $A^{\ast}$ as a function of contact angles
  $(\phi_1,\phi_2)$ that enclose regions of integrity across the
  whole shear stress plateau. On the dashed line $\phi_1=\phi_2$. For
  $A^*=-3.5$ the dotted arrows indicates the contact angles for
  which the meniscus will be integral for all $\hat{w}_2$. The dotted lines indicate fracture at some $\hat{w}_2$ as shown in Fig.~\ref{fig:Fig5}a.}
\label{fig:Fig3}
\end{figure*}
\begin{table*}[htb]
\begin{tabular}{rcl@{\hspace{12truept}}l}
  \hline\hline
  \multicolumn{3}{c}{Range of $A$}& Integrity of meniscus\\\hline
  & $A$ & $<-4$ &  {\footnotesize Fractures for all contact angles
    $\phi$, for $\hat{w}^-_2 < \hat{w}_2 < \hat{w}^+_2$}.\\ 
  $-4<$ & $A$ & $<-2$ & \footnotesize Fractures for 
    $|\cos\phi| >  |\cos\phi^{\ast}|$ and 
 $\hat{w}^-_2 < \hat{w}_2<\hat{w}^+_2$.
\\  
  $-2<$ & $A$ & $<\;\;\:0$ & {\footnotesize Integral for all
    contact angles $\phi$.}\\
  \hline\hline
\end{tabular}
\caption{Criteria for (lack of) integrity as a function of the distortion
  parameter  $A$, for equal contact angles $\phi_1=\phi_2=\phi$.} 
\label{tab:one}
\end{table*}
\subsection{Different Contact angles}
For different contact angles the situation is more complex. For $A<0$
the meniscus remains integral for
all $\hat{w}_2$ for $A>-8$ and 
\begin{displaymath}
\cos\phi_2>\left[\sqrt{|\cos\phi_1-1|}-\sqrt{|A|}\right]^2-1.
\end{displaymath}
For $A<-8$ the meniscus fractures somewhere along the stress plateau
for any combination of contact angles. 
This condition is illustrated in Fig.~\ref{fig:Fig3}, which shows the  
regions of integrity and fracture as a function of both contact
angles, for $A=-3.5$ (Fig.~\ref{fig:Fig5}a is a slice through this
figure for $\phi_1=\phi_2$). Fig.~\ref{fig:Fig3}b shows that an
asymmetry in contact angle increases the region for meniscus fracture. For
 $\phi_1=0$, corresponding to a wetting surface, a larger contact
 angle leads to a fractured meniscus for $|A|<2$, which would always be integral for
 equal contact angles. 

\subsection{Pinned Contact Lines}
\begin{figure*}[htb]
\begin{center}
\includegraphics[width = 1\textwidth]{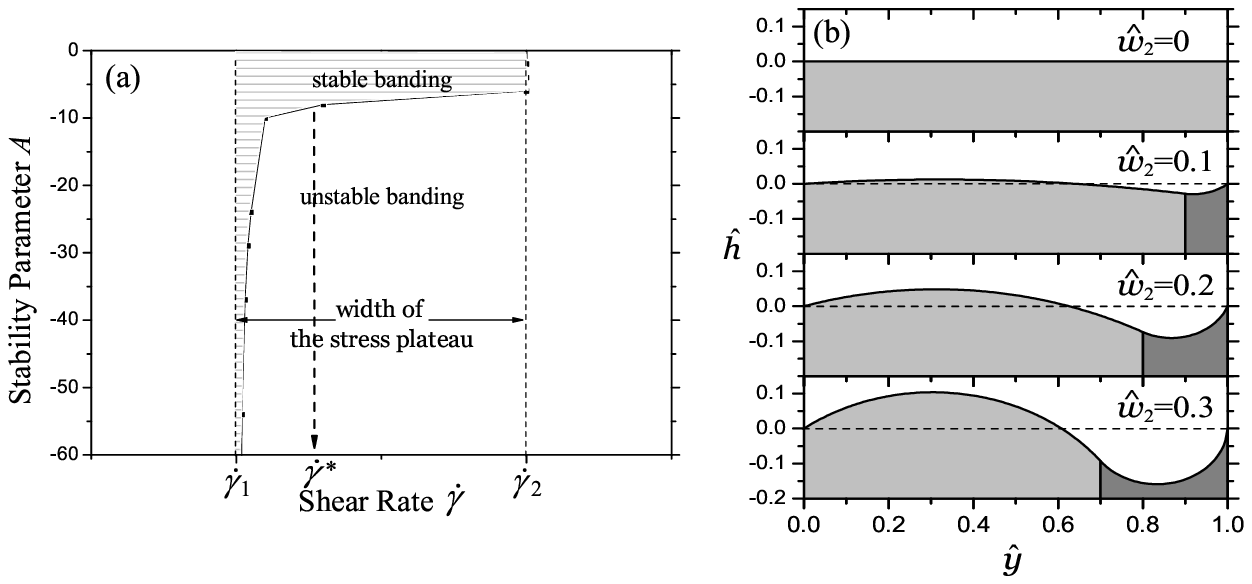}
\end{center}
\caption{Stability regions of profiles for contact lines pinned such that $H=0$. (a) The full width of the shear stress plateau is accessible for $-8<A<0$. For $A<-8$ the surface will fracture at some applied shear rate $<\dot{\gamma}^\ast$. (b) Meniscus profiles for $A=-8$. The surface maintains integrity for applied shear rates up to $\dot{\gamma}^\ast$ at which rate the high shear rate band has achieved $w_2=0.3$ of the width of the gap.}
\label{fig:Pinned}
\end{figure*}
For pinned contact lines we ascribe a particular value to $H$. Eqs~\eqref{eqn:deltaN}, \eqref{eqn:topC} and ~\eqref{eqn:topA} are not sufficient to specify $R_1,\;R_2,\;\phi_1$ and $\phi_2$; the fourth condition is the requirement of volume conservation for an incompressible fluid. Fig.~\ref{fig:Pinned} is a typical example in which the contact lines have been pinned so that their difference in height $H$ is zero. For applied shear rates associated with the stress plateau the meniscus contorts but retains integrity for all shear rates when $-8<A<0$; for $A<-8$ not all shear rates are accessible and the meniscus will fracture at some shear rate.

\subsection{Three Bands}
Fig.~\ref{fig:Cells} shows an example of three shear bands, visualized
in a cone-and-plate rheometer by \citet{BritCall97c} using NMR
velocimetry. Three bands have not been observed in cylindrical Couette
flow, and this difference was rationalized by \citet{BCAdams08} as due
to a combination of boundary conditions that favor the low shear rate
phase and the relatively weak stress gradient of cone-and-plate flow.
Motivated by this result, we have calculated the distortion of the
meniscus in such a configuration and find that a three band state does indeed allow meniscus integrity, for certain ranges of parameters. This analysis is given in Appendix~\ref{sec:three-band-conf}. 
\section{Comparison with the literature}\label{sec:application-model}
\subsection{Theoretical Method} 
We examine existing data in the literature,
which exhibits both stable shear banding across the entire stress
plateau and an instability such that the entire stress
plateau could not be accessed; we then assess whether or not
meniscus fracture is expected, based on an estimation of the
distortion parameter $A=W\Delta N_2/\gamma_s$.  We fit the data from
the flow curves, including the stress plateau and the shear rates
$\dot{\gamma}_1$ and $\dot{\gamma}_2$ in the two shear bands, to both
the Giesekus and Johnson-Segalman (JS) constitutive models. From the models
we can evaluate the predicted second normal stress difference $N_2$ in
the two shear bands (see Appendix~\ref{sec:const-equat}), while the
gap size $W$ is taken from the experimental conditions and the surface
tension $\gamma_s$ is estimated using literature values.  With this in
hand we can then compare the stability or instability of banding
states, inferred from the experiments, with our calculations. Since
the contact angles are unknown, we can only determine whether our
criteria for instability are consistent with the numerical values for
$A$ (assuming equal contact angles). We are also limited by the quality of the available constitutive models: neither the JS or Giesekus models are expected  specifically apply to wormlike micelles, but they can support shear banding and have non-zero second normal stresses. These models, developed for polymer melts, may be more applicable to semi-dilute polymeric solutions, which have an explicit Newtonian solvent. \citet{FiscReha97}, \citet{yesilata2006nonlinear}, and \citet{Helg} show that the steady state non-linear rheology and shear thinning of worm-like micelles is well described by the Giesekus model; we offer the Johnson-Segalman model for comparison. 
In fact, the two models yield very similar quantitative predictions and thus do not substantially differ insofar as determining the properties of the meniscus.
\subsection{Wormlike Micellar Solutions}
\begin{figure*}[htb]
\includegraphics[width = 1\textwidth]{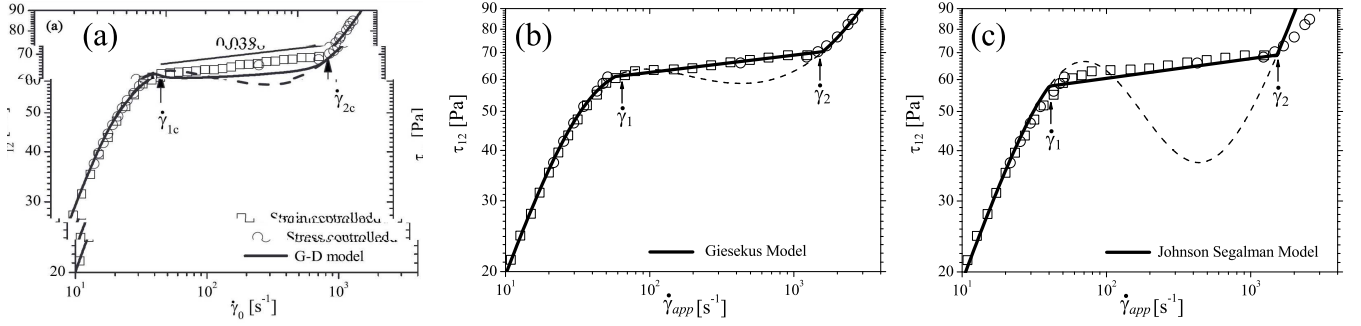}
\caption{(a) Flow curve for a CTAB solution fitted to the (numerical)
  diffuse Giesekus model, measured in a Couette rheometer with a gap
 $W=0.5\,\textrm{mm}$ \citep{Helg}. Fits to the non-diffuse
  Giesekus (b) and JS (c) models (fitting parameters are
  given in Table \ref{tab:six}).}
\label{fig:Helg}
\end{figure*}

Wormlike micellar solutions of both cetyltrimethylammonium (CTAB) and
cetylpyridinium choloride/sodium salicilate (CPCl/NaSal) have been
extensively studied.   
Fig.~\ref{fig:Helg} shows the data of a CTAB solution, as measured by
\citet{Helg} and fit by them to the diffuse Giesekus model. In this case the entire stress plateau was accessible. We have also fitted it to the non-diffuse Giesekus and JS models.
Strictly, one should fit the stress plateau using a non-local (or
diffuse) model \citep{lu99}; however, our fits obtained by choosing
the stress plateau `by hand' differ insignificantly from more precise
fitting. Hence we use a local model for the remaining fits in this
paper. The Giesekus model fits the high shear rate branch much better than does the JS model. 

The fitting parameters are shown in Table~\ref{tab:six}. The
difference between the diffuse and non-diffuse Giesekus models values
for $\Delta N_2$ is less than 3\%. We estimate the distortion parameter
to be $A\simeq-0.8$, which is well within range ($A\geq-2$, from
Table~\ref{tab:one}) for which we expect to find stable shear bands
across the entire stress plateau for all contact angles.

\begin{table*}[htb]\centering
\begin{tabular}{cc@{\hskip4.0truept}c@{\hskip8.0truept}c@{\hskip8.0truept}c@{\hskip8.0truept}ccccccccl}
\hline\hline
& & & & & \multicolumn{8}{c}{Parameters} \\
 \cline{6-13}
Figure & Material & $\eta_o/(\textrm{Pa\,s})$ & $\text{T}/^\circ
\text{C}$ &$W/\textrm{mm}$&  $G/\textrm{Pa}$ & $\tau/\textrm{s}$ & $a$ &
$\epsilon$ & $\dot{\gamma}_1 / \textrm{s}^{-1}$ &
$\dot{\gamma}_2/\text{s}^{-1}$ & $\frac{\Delta N_2}{G}$ &  $A$ &
\\ 
\hline
Fig.~\ref{fig:Helg} & CTAB &  2.1 & 32  &  0.5&112 & 0.018 & 0.9
& 0.008 & 65 & 1530 & -0.48 &   -0.8 &  Gi \\
& & & & & 115 & 0.017 & 0.43 & 0.022 & 40 & 1430 & -0.52 & 
-0.8 &  JS \\ [1ex] 
Fig.~\ref{fig:Berret(8pc)}&  CPCl & 18 & 30 & 0.5 & 165 & 0.11 &
0.88 & 0.015 & 11 & 60 & -0.38 &  -0.9 &   Gi \\
&  8\% & & & &168 & 0.09 & 0.41 & 0.056 & 8 & 90 & -0.48 &  
-1.2 &  JS\\ [1ex] 
Fig.~\ref{fig:Berret(12pc30)} & CPCl & 240 & 20  & 0.5& 260 &
0.94 & 0.87 & 0.002 & 1 & 54* & -0.60 &  -2.2 &  Gi \\
& 12\% & & & &268 & 0.88 & 0.38 & 0.010 & 0.85 & 54* & -0.49 & -1.9 &  JS\\
\hline\hline 
\end{tabular}
\caption{Parameters used to fit the flow curves in
  Figs.~\ref{fig:Helg}, \ref{fig:Berret(8pc)}, and
  \ref{fig:Berret(12pc30)}, for the Giesekus (Gi) or Johnson-Segalman (JS) models.   We estimate a surface tension
  $\gamma_s\simeq37\,\text{mN/m}$ for aqueous CTAB
  \citep{Christian}  and $\gamma_s\simeq32\:\text{mN/m}$ for CPCl/NaSal
  \citep{akers2006impact}. For the stable solutions (CTAB and 8\% CPCl) the shear rate $\dot{\gamma}_2$ of the high shear rate band was taken from the end of the stress plateau, while for the unstable solution (12\% CPCl) $\dot{\gamma}_2$ was estimated from the scaling in Fig.~8 of \citet{BPD97}.
}
\label{tab:six}
\end{table*}

We have found a few experiments that show a clear instability as the
stress plateau is traversed, or that cannot reach the end of the
stress plateau.  Berret and co-workers (\citet{BRP94} and \citet{BPD97}) studied CPCl/NaSal
solutions at different concentrations and temperatures, in a
cone-and-plate rheometer with a free surface. They found stable stress
plateaus at micellar solutions close to a non-equilibrium critical
point, where the difference between the shear banding phases vanishes.
Farther from the critical point, where the shear banding phases are
more distinct, and hence one expects $\Delta N_2$ to be larger, they
report {unstable stress plateaus (see Fig.~6 of \citet{BPD97}).
Fig.~\ref{fig:Berret(8pc)} show the data and fits for an $8\%$
solution at $T=30^{\circ}\,\textrm{C}$, which was close to the non-equilibrium critical point. The fitting leads to a distortion parameter between $A\simeq-1.0$ and $A\simeq-1.3$, which is consistent with
the meniscus maintaining integrity across the entire stress plateau.

The data for an unstable CPCl solution are shown in
Fig.~\ref{fig:Berret(12pc30)}, at $12\%$ and
$T=20^{\circ}\,\textrm{C}$. In this case the stress plateau can only be traversed
as far as $\dot{\gamma}_{\textrm{app}}=7.1\,\textrm{s}^{-1}$, at which
point the fluid became unstable \citep{BPD97}. To calculate the
distortion parameter we require the shear rate $\dot{\gamma}_2$ in the
high shear rate phase, which we estimate from Fig.~8 of \citep{BPD97}
to be $\dot{\gamma}_2\simeq54\,\textrm{s}^{-1}$. This is consistent
with the measurements of the high shear rate branch performed by
\citet{LopGon} on the same material with a free surface: Fig.~11 of
their paper shows a a $10\%$ sample at $25^{\circ}\,\textrm{C}$. Fits to the 
Giesekus and JS models respectively yield $A\simeq-2.4$ and
 $A\simeq-2.0$, for the $12\%$ sample of \citet{BPD97}, at $T=20^{\circ}\,\textrm{C}$. 
In the range $A<-2$ we expect meniscus fracture for
some contact angles, and across some region of the stress plateau. The
experiments show an instability at
$\hat{w}_2=0.12$. Fig.~\ref{fig:Fig6} shows that such a small value of
$\hat{w}_2=0.12$ and $A\alt-2$ would be consistent with a contact
angle near $180^{\circ}$, or complete wetting. We caution, however,
against using the numerical results from applying these fairly crude
constitutive models. It is clear, however, that the distortion
parameter predicts that the 12$\%$ fluid at $T=20^{\circ}\,\textrm{C}$ should be
more unstable than the $8\%$ fluid at $T=30^{\circ}\,\textrm{C}$. We have also
estimated the {distortion parameter from the closely related experiments
of \citet{LopGon} (see Fig.~5 from that paper), for which we estimate
$-A\simeq2.3-2.6$, again just into the predicted range of possible meniscus fracture.

\begin{figure*}[htb]
\includegraphics[width = 1\textwidth]{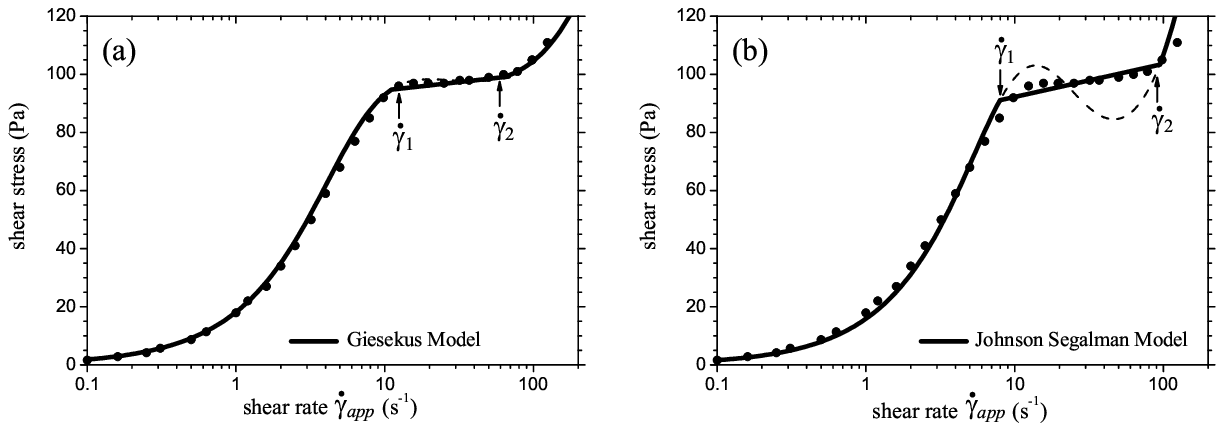}
\caption{Data and fitted constitutive curves for the experiments of
  \citet{BPD97} on CPCl at $8\%$ (by weight) and $T=30^{\circ}\,\textrm{C}$, using the (a)
  Giesekus (b) JS models. The geometry was a cone and
  plate rheometer. Parameters are shown in Table~\ref{tab:six}.}
\label{fig:Berret(8pc)}
\end{figure*}
\begin{figure*}[htb]
\includegraphics[width = 1\textwidth]{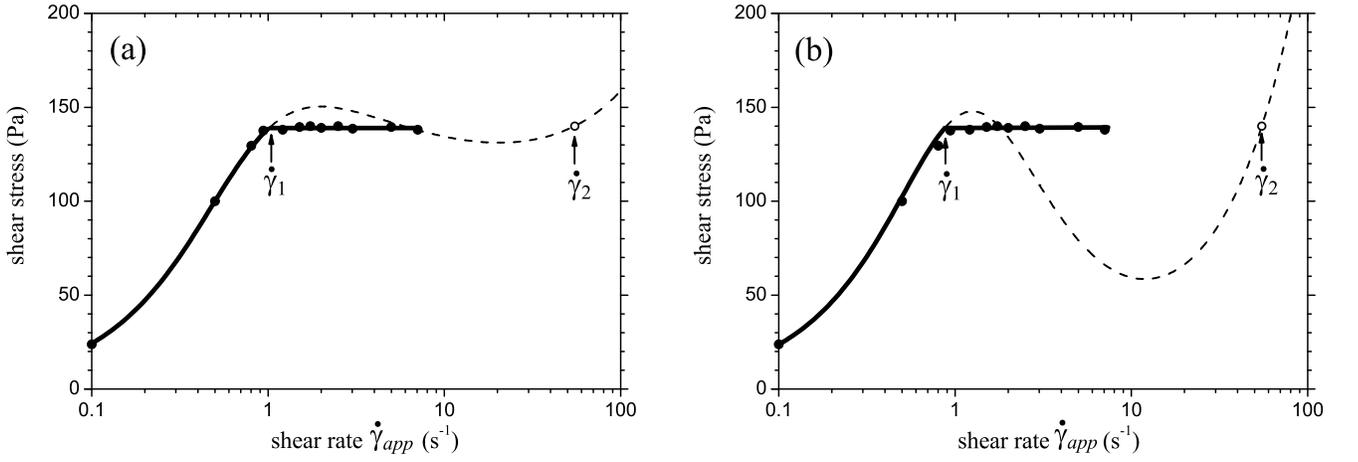}
\caption{Data and fitted constitutive
  curves for the experiments of \citet{BPD97} on CPCl at $12\%$ (by weight) and
  $T=20^{\circ}\,\textrm{C}$, using the (a) Giesekus (b) JS
  models. The geometry was a cone and plate rheometer. Parameters are
  shown in Table~\ref{tab:six}. The fluid became unstable at the
  rightmost data point $\blacksquare$, at
  $\dot{\gamma}_{\textrm{app}}=7.2\,\textrm{s}^{-1}$. The open circle $\circ$
  shows our estimate of the shear rate in the high shear rate branch based on extrapolation from the rheology of Fig.~8 of \citet{BPD97}, together with the temperature dependences in Figs.~5 and~7 of the same reference and Fig.~4 of \citet{BRP94} .
  }
\label{fig:Berret(12pc30)}
\end{figure*}

We have collected much of the available data in the literature in
Table~\ref{tab:two}. In many cases the full banding
profile was found (noted as `stable' in Table~\ref{tab:two}), and in
all of these cases we calculated $A$ to be in the range $A>-2$, the meniscus retains integrity
(Table~\ref{tab:one}). Similarly, all of the unstable data that we
have found corresponds to $A<-2$, for which meniscus fracture is expected
for some contact angles (Table~\ref{tab:one}).

\begingroup
\squeezetable
\begin{table*}[htb!]\centering
\begin{tabular}{c@{\hskip4.0truept}cc@{\hskip4.0truept}c@{\hskip4.0truept}cccccccccccc}
\hline\hline
& & & \multicolumn{8}{c}{Model Parameters}& & & & \multicolumn{2}{c}{Instability at}\\
 \cline{5-12}\cline{15-16}
Material & Cell & $\text{T}/^\circ
\text{C}$ &$W/\textrm{mm}$&  $G/\textrm{Pa}$ & $\tau/\textrm{s}$ & $a$ &
$\epsilon$ & $\dot{\gamma}_1 / \textrm{s}^{-1}$ &
$\dot{\gamma}_2/\text{s}^{-1}$ & $\frac{\Delta N_2}{G}$ & $A$ & & &$\dot{\gamma}_{\textrm{app}}^{\ast}/\textrm{s}^{-1}$& $\hat{w}_2^{\ast}$ 
\\ 
\hline
 CTAB\footnotemark[1]& C & 32 & 0.17& 69 & 0.11 & 0.92 & 0.005 & 6 & 330 & -0.71& -0.2 & Gi & stable & & \\
 & & & & 70 & 0.11 & 0.5 & 0.01 & 5.5 & 450 & -0.53 &  -0.2& JS & stable & & \\[1ex]
CPCl\footnotemark[2] & CP&  20 & 0.5& 94& 0.62 & 0.92 & 0.02 & 1.5 & 11& -0.48& -0.7& Gi & stable & & \\
 6\% & & & & 96 & 0.55 & 0.38 & 0.055 & 1.2 & 13 & -0.51 & -0.8 & JS & stable & & \\[1ex] 
CTAB\footnotemark[3] & C & 32  &  0.5&112 & 0.018 & 0.9
& 0.008 & 65 & 1530 & -0.48 &  -0.7 &  Gi & stable & & \\
  Fig.~\ref{fig:Helg}&& & & 115 & 0.017 & 0.43 & 0.022 & 40 & 1430 & -0.52 & 
 -0.8 &  JS & stable & & \\ [1ex] 
 CPCl\footnotemark[2] & CP &  30 & 0.5 & 165 & 0.11 &
0.88 & 0.015 & 11 & 60 & -0.38 &  -1.0 &   Gi & stable & & \\
 8\% Fig.~\ref{fig:Berret(8pc)} & & & &168 & 0.09 & 0.41 & 0.056 & 8 & 90 & -0.48 &  
 -1.3 &  JS & stable & & \\ [1ex] 
 CTAB\footnotemark[1]  & C &  32 & 1 & 69 & 0.11 & 0.92 & 0.005 & 6 & 330 & -0.71 & -1.3 & Gi & stable & & \\
 & & & & 70 & 0.11 & 0.5 & 0.01 & 5.5 & 450 & -0.53 & -1.0 & JS & stable & & \\[1ex]
CPCl\footnotemark[4] & C & 21.5 & 1& 108 & 0.48 & 0.88 & 0.012 & 2.5 & 26 & -0.44 & -1.5 & Gi & stable & & \\
  6\% & & & & 105 & 0.51 & 0.65 & 0.036 & 1.5 & 27 & -0.45 & -1.5 & JS & stable & & \\[1ex] 
 CPCl\footnotemark[2] & CP & 30 & 0.5& 291 &  0.15 & 0.84 & 0.002 & 8 & *187 & -0.50 & -2.3 & Gi & unstable & 80 &  0.40 \\
 12\%& & & & 282 & 0.14 & 0.57 & 0.021 & 6.5 & *187 & -0.42 & -1.8 & JS & unstable & & 0.41\\[1ex]
CPCl\footnotemark[2] & CP & 20  & 0.5 & 260 &
0.94 & 0.87 & 0.002 & 1 & *54 & -0.59 &  -2.4 &  Gi& unstable  & 7.1 & 0.12\\
 12\% Fig.~\ref{fig:Berret(12pc30)} & & & &268 & 0.88 & 0.38 & 0.010 & 0.85 & *54 & -0.49 &  -2.0 &  JS & unstable & & 0.12\\[1ex]
CPCl\footnotemark[5] & CP & 25 & 0.6& 224 & 0.32 & 0.92 & 0.007 & 2.5 & *70 & -0.62 & -2.6 & Gi & unstable & 43 & 0.60\\
 10\% & & & & 225 & 0.29 & 0.4 & 0.022 & 2.2 & *70 & -0.54 & -2.3 & JS & unstable & & 0.60\\[1ex]
 CTAB\footnotemark[6] & C & 28 & 1.13& 267 & 0.21 & 0.89 & 0.016 & 4 & 90 & -0.59 & -4.8 & Gi & *unstable & (40)& (0.42)\\
 & & & & 290 & 0.18 & 0.43 & 0.033 & 3 & 95 & -0.56 & -5.0 & JS & *unstable & & 0.40 \\ 
\hline\hline 
\end{tabular}
\footnotetext[1]{\citet*{Capp+97b}.}
\footnotetext[2]{\baselineskip=12truept\citet*{BPD97}. The shear rate in the high shear rate branch for the 12\% solution was estimated by extrapolating from Fig.~8 in this reference. }
\footnotetext[3]{\citet*{Helg}.}
\footnotetext[4]{\citet*{SalmonPRL90}.}
\footnotetext[5]{\baselineskip=12truept\citet*{LopGon}. The shear rate in the high shear rate branch for the 10\% solution  was estimated by extrapolating from Fig.~8 in \citet{BPD97}.}
\footnotetext[6]{\baselineskip=12truept\citet*{lerouge2008interface}. Although the data in
  this reference are apparently stable because the entire stress plateau
  is traversed,  the cell top was covered: for an open cell with a free surface  an instability
  occurs at 
  $\dot{\gamma}_{\textrm{app}}^{\ast}\simeq 40\,\textrm{s}^{-1}$  (Lerouge, private communication).}
\caption{Values of the distortion parameter $A$ calculated
  from experimental studies on worm like micellar solutions using
  either the Giesekus (Gi) or JS
  models. ``Stable''  experiments  accessed
  the full stress plateau, while  in ``unstable'' cases  an instability
  occurred at the apparent shear rate $\dot{\gamma}_{\textrm app
  }^{\ast}$  for a high shear rate band of width $\hat{w}_2^{\ast}$. 
 The first column identifies the
  figure above that demonstrates the fit, and the
  weight fraction used for the  experiments of
  \citet{BPD97}. We estimate a surface tension
  $\gamma_s\simeq37\,\text{mN/m}$ for aqueous CTAB
  \citep{Christian}  and $\gamma_s\simeq32\:\text{mN/m}$ for CPCl/NaSal
  \citep{akers2006impact}.}
\label{tab:two}
\end{table*}
\endgroup

\subsection{Polymer Solutions}
Wang and co-workers revived the experimental study of entangled
polymers with new experiments that clearly show shear banding
\citep{tapadiawang03,hu2007cre,boukany07,boukany2009shear}.
\citeauthor{tapadiawang03}'s first experiments, on polybutadiene (PBD)
entangled in its own oligomer, suggested shear banding by what they
referred to as an `entanglement disentanglement transition' (EDT);
this data is also broadly consistent with theories based on the
Doi-Edwards tube model
\citep{adamsolmsted09a,adamsolmsted09b,wangcomment2009}. In the earliest
experiments they apparently used a free surface, and commented that
some small edge fracture or instability could have occurred, but that
its effects were negligible.

While  attempting to reproduce these experiments, \citet{inn2005eef} found
significant edge fracture and instability in the region of the
transition and the stress plateau, and concluded that the surface
instability and associated mass loss was responsible for the effects.
This was then studied by \citet{sui2007iep} using cone-and-plate
rheometers with gap sizes at the rim of $W=0.267\,\textrm{mm}$ and
$W=1.363\,\textrm{mm}$. They found significant edge fracture or
instability accompanying the shear banding transition, but their
results suggested that the surface instability could be a consequence,
rather than a cause, of the shear banding transition (or EDT). This is
seen most clearly in Fig.~5 of \citet{sui2007iep}, which shows that
the stress plateau can be traversed when a plastic film is used to
suppress the instability at the surface (Philips and Wang performed
the film-suppressed experiments). \citet{schweizer2007sbd} also found
that surface deformation accompanied the banding, or EDT, in similar
materials. More recently, \citet{ravindranath2008steady} and
\citet{liwang2010yield} have verified that the shear banding (or EDT)
transition can persist when even when surface effects are suppressed.

\begin{figure*}[!htb]
\includegraphics[width = 1\textwidth]{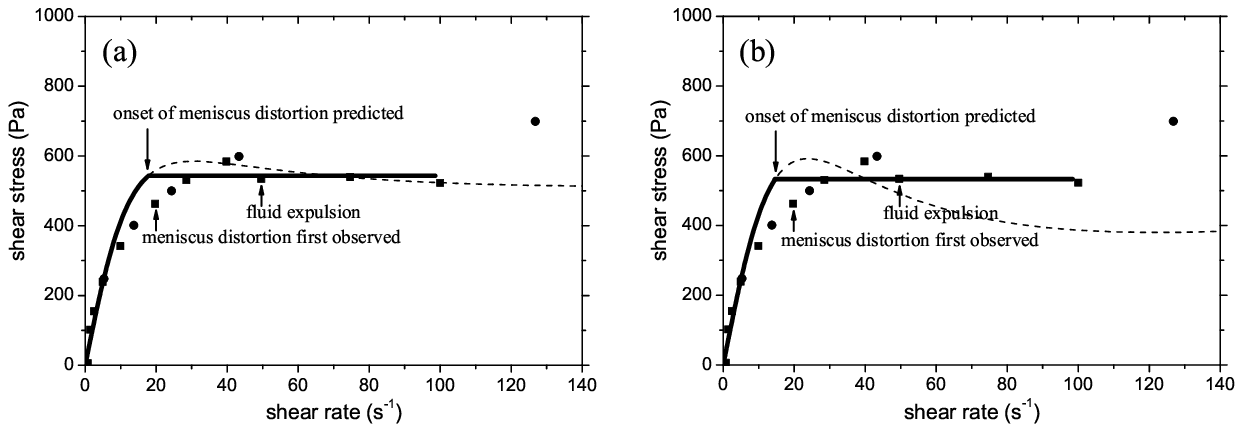}
\caption{Data and flow curves for the experiments of
  \citet{sui2007iep} on SRM 2400, using a $6^{\circ}$ cone
  angle. $\blacksquare$ indicates rate-controlled and {\LARGE
    \textbullet} stress-controlled rheometry. Fits are to the (a) Giesekus and (b)
  Johnson Segalman models.}
\label{fig:SRM}
\end{figure*}
\begin{figure*}[!htb]
\begin{center}
\includegraphics[width =1\textwidth]{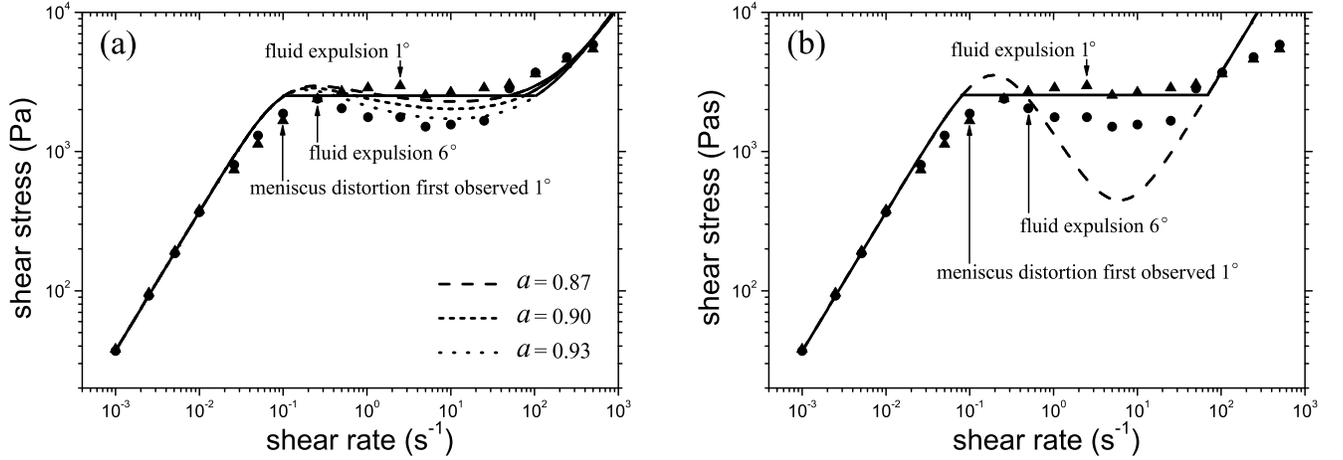}
\end{center}
\caption{Data and flow curves for the experiments of
  \citet{sui2007iep} on PBD  under controlled strain rate conditions, fitted to either the Giesekus (a) or JS (b) models. The Giesekus fits show several different fits, to illustrate the difficulty of precise characterization of the constitutive models. Table \ref{tab:seven} shows that the distortion parameter $A$ is only very weakly sensitive to these different fits. Here, $\blacktriangle$ indicates $1^\circ$ gap and {\LARGE \textbullet} indicates $6^\circ$.}
\label{fig:PBD}
\end{figure*}
\begingroup
\squeezetable
\begin{table*}[!htb]\centering
\begin{tabular}{cc@{\hskip4truept}cc@{\hskip8truept}c@{\hskip8truept}ccccccccl@{\hskip8truept}cc}
\hline\hline
 &  &cone& & &\multicolumn{8}{c}{Parameters} & & \multicolumn{2}{c}{Instability at}\\
 \cline{6-13}\cline{15-16}
Material & $\eta_o/(\textrm{Pa\,s})$ &angle& $\text{T}/^\circ\text{C}$&$W/\textrm{mm}$ &  $G/\text{Pa}$ &
$\tau/\text{s}$ & $a$ & $\epsilon$ 
& $\dot{\gamma}_1/\text{s}^{-1}$ &
$\dot{\gamma}_2/\text{s}^{-1}$ & $\frac{\Delta N_2}{G}$ &
  $A$ & & $\dot{\gamma}_{\textrm{app}}/\textrm{s}^{-1}$& $\hat{w}_2$ \\ 
\hline
SRM 2490 & $85$ & 1$^\circ$ & \text room & 0.267& 1060 & 0.047 & 0.93 & 0.01 &
18 & 300* & -0.571 &  -6.2 &   Gi & \text none \!\!\! & \\
& & & ($\sim 24$)& 0.267 & 1040 & 0.048 & 0.36 & 0.03 & 17 & 300* & -0.461 & -4.9 & JS & noted & \\[1ex]
 &  & 6$^\circ$&  & 1.363& 1060 & 0.047 & 0.93 & 0.01 & 18 & 300* &
 -0.571 & -32&  Gi & 50 & 0.11\\
 &  & & &1.363&  1040 & 0.048 & 0.36 & 0.03 & 17 & 300* & -0.461 & -25 &
  JS & & 0.12\\  
  \hline
PBD & $3.9\times 10^4$ & 1$^\circ$ & 30 & 0.267& 5150 & 7.2 & 0.87 & 0.0001 &
0.1 & 50 & -0.717 &  -33 &   Gi & 2.5 & 0.05\\
 & & & & &'' &'' & 0.90 &  '' &'' & 71 & -0.709 & -32 &Gi & ''&''\\
 & & & & &'' &'' & 0.93 &  '' &'' & 103 & -0.697 & -32 & Gi&''&'' \\ 
 &  & & & & 5170 & 7.1 & 0.68 & 0.001 & 0.08 & 60 & -0.515 & -24 &
  JS &''&''\\ [1ex] 
 &  & 6$^\circ$&  & 1.363& 5150 & 7.2 & 0.87 & 0.0001 & 0.1 & 50 &
 -0.717 & -167&  Gi & 0.25 & 0.00\\
& & & & & '' &'' & 0.90 & '' &'' & 71 & -0.709 & -166 & Gi& ''&''\\
 & & & & & '' &'' & 0.93 &  '' &'' & 103 & -0.697 & -163 &Gi &''&'' \\  
 &  & & &1.363&  5170 & 7.1 & 0.68 & 0.001 & 0.08 & 50 & -0.515 & -121 &
  JS & ''&''\\ 
\hline\hline 
\end{tabular}
\caption{Parameters used to fit the flow curves in Figure
  \ref{fig:SRM}a and \ref{fig:SRM}b  to the Johnson-Segalman (JS) and Giesekus (Gi) and evaluate the distortion
  parameter $A$ for the study of \citet{sui2007iep} of
  polybutadiene-in-oligomer 
  (PBD) and polyisobutylene-in-pristane (SRM 2490)
  solutions.  The
  surface tension of pristane  is around $26\,\text{mN/m}$ 
  \citep{pristane} and that of 1,4 PBD is around
  $30\,\text{mN/m}$ \citep{polymerhandbook}.  The PBD was fit to the Giesekus model for several possibilities to illustrate the robustness of the parameter $A$. $^{\ast}$For the SRM solutions the shear rate $\dot{\gamma}_2$ of the high shear rate band was estimated based on the parametrization of the shear thinning behaviour of the low shear rate branch and  the stress plateau. The parameter $A$  is very insensitive to the precise value of $\dot{\gamma}_2$:  values of $\dot{\gamma}_2$ from 200 to 600\,$\textrm{s}^{-1}$ change $A$ by less than $10\%$.}
\label{tab:seven}
\end{table*}
\endgroup
To address these issues we have estimated the distortion parameter for
the experiments of \citet{sui2007iep} on polybutadiene solutions, and
on solutions of polyisobutylene in pristane (SRM 2490) (Table \ref{tab:seven}).
The distortion parameter for PBD is quite large and negative, between
$A=-120$ and $A=-20$ for these experiments,
such that instability should be present early into the stress plateau
for all contact angles. This is consistent with the experimental
results noted above: shear banding can be observed if the meniscus is
sealed, and an instability occurs otherwise. 

On the other hand, we estimate the lower viscosity material SRM2490 to have a much smaller $A$. These values are consistent with the data of
\citet{sui2007iep}, who noted explicitly
that reducing the rim gap delays the onset of instability and
expulsion as does reducing the modulus ($G$); both of these
changes reduce $|A|$. They  report  controlled strain rate data  for the larger $6^{\circ}$ cone angle (and wider gap), for which we estimate $A\simeq-30$, which is well within the regime where one expects fracture; see Fig.~\ref{fig:SRM}. They observe noticeable meniscus distortion at $\dot{\gamma}\simeq20\,\textrm{s}^{-1}$, and fluid expulsion at
$\dot{\gamma}\simeq50\,\textrm{s}^{-1}$. This suggests only a very small window of stability on the stress plateau, which is consistent with our calculations.
For the smaller $1^{\circ}$ cone angle they only perform controlled stress experiments, and did not observe mass loss at the accessible strain rates ($\sim95\,\textrm{s}^{-1}$). We estimate values of $A\simeq-4.9$ (JS) or $A=-6.2$ (Giesekus), which is just beyond the limits where we predict meniscus integrity across the shear plateau. Hence, their controlled stress experiments can access a large part of the stress plateau; moreover,  we expect that an fracture would occur before the end of the plateau is reached, based on earlier work on the same fluid  by \citet{NISTSRM}. In their discussion they noted:
\begin{quote}
\textsl{
``The conditions at the edge of the cone and plate [gap $0.49\,\text{mm}$] can impact the measurements in several ways, but these effects are not easily quantifiable. Perhaps the most significant difficulty is that the
fluid can escape from between the cone and plate. One indicator of loss of fluid would be a
decrease in the moment with increasing shear rate. This decrease was only observed in three
experiments at $0^\circ\text{C}$ in the step from a shear rate of $63\,\text{s}^{-1}$ to a shear rate of $100\,\text{s}^{-1}$. Those three measurements were discarded. The only other evidence of edge effects occurs at the three highest shear rates at all three temperatures [0, 25 and 50$^\circ\text{C}$], where there is an increase in the relative scatter of the viscosity data. For this reason, data at the three highest shear rates [40, 63 and $100\,\text{s}^{-1}$] are provided as reference
data only, since the sample geometry might not match our assumptions, and the uncertainty in the
data cannot be completely quantified.'' --- \citep[p.~22]{NISTSRM}}
\end{quote}
We estimate the stress plateau to end at $\dot{\gamma}\approx 300\:{\rm s}^{-1}$. The measurements of \citet{NISTSRM}, with a cone and plate gap of $0.49\;{\rm mm}$, thus show that meniscus distortion has started by $40\:{\rm s}^{-1}$. Even allowing for the narrower gap of the \citet{sui2007iep} $1^\circ$ cone and plate, we would expect their fluid to fracture before achieving a shear rate of $300\: {\rm s}^{-1}$.   

\section{Conclusion}\label{sec:conclusion}
\subsection{Discussion}
Our meniscus distortion calculation leads to a result that is similar to that of
\citet{Keentok}, which was not devised for shear banding fluids. They
considered a more detailed calculation that incorporated the flow
field due to a small perturbation, but also led to a stability
condition that balanced the normal stresses with surface tension. By
contrast, we have effectively assumed that the radius of curvature is
large enough, compared to the gap size, so that one can focus entirely
on the integrity of the free surface. Hence, we expect our condition
for fracture of the meniscus to be preempted by secondary flows as the radius of
curvature decreases.

It would be difficult to experimentally probe 
the dependence on
contact angle, because entirely different sample cells or surface
preparations would be needed. However, for highly viscous materials
the contact angles can be set by the loading protocol
\citep{Schweizer,schweizer2007sbd}, and experiments performed much more
quickly than the true equilibrium contact angle can be reached. Hence,
one could systematically vary the contact angle to qualitatively test
our predictions (\textit{e.g.} Fig.~\ref{fig:Fig3}). Other more obvious experimental tests would be to change the gap size and directly observe the deformation of the meniscus as shear banding proceeds. 

Although our calculation was performed with shear banding fluids in mind, our predictions could be tested on simpler fluids. For example, two immiscible fluids with different second normal stress behaviors could be prepared as discs of different thicknesses below $T_g$, loaded into a cone and plate rheometer, and then brought into the melt state before shearing and observing the meniscus.
\subsection{Summary}
Motivated by edge-fracture-like instabilities that occur during shear
banding, we have calculated the distortion of the free meniscus in the
gradient shear banding configuration. The second normal stress
difference between two shear bands determines the radii of curvature
of each band, and the conditions of continuity and smoothness of
the interface lead to integrity conditions for the interface. The
integrity limit is defined to occur when the meniscus is entirely
vertical and thus on the point of overhanging itself; this is probably a conservative estimate of meniscus integrity, since we have not explicitly calculate the detailed velocity field in the region of the surface. 
It must be stressed that we have presented a simplified calculation ignoring the complicated flow field and shear conditions near the surface. The necessary calculation has been outlined in Appendix \ref{sec:free-surface}. 

We have calculated
integrity diagrams as a function of contact angle and the average
applied shear rate, for given values of the distortion parameter
$A=W\Delta N_2/\gamma_s$. Since the sample will climb up one wall to
to maintain integrity, expulsion would usually be expected to occur
before our theoretical integrity limit has been reached. The integrity
diagrams depend only on the values of $A$ and the contact angles,
regardless of the constitutive behaviour. 

By comparison with parameters extracted from experiments, we find that wormlike micellar
solutions are often expected to have stable shear bands, although they
can sometimes be unstable; this is consistent with the experimental record.  \citet{BPD97} found that micellar
solutions became more unstable when the shear bands were more
different from each other (\textit{i.e.} farther from the
non-equilibrium critical point in concentration-temperature space),
and at lower temperatures where the modulus, and hence $N_2$ and thus $A$, are
higher.

Entangled polymer solutions, on the other hand, are predicted to
easily become unstable due to their much higher normal stresses. This
is consistent with the recent body of experiments that show shear
banding in entangled polymers: unless the surfaces are protected or
the meniscus shielded, the meniscus develops an instability
\citep{inn2005eef,sui2007iep}.

We  also considered the {meniscus integrity of three bands, as was seen by \citet{Britton.Callaghan99} in a cone and plate rheometer. [A cone and plate is a good candidate for three bands because of its the very weak stress gradient due to curvature \citep{BCAdams08}.] To resolve whether or
not the three bands can be symmetrically arranged, we calculated the
surface energy (proportional to the surface tension). The stability of
this energy leads to a rich variety of possible configurations: the
high shear rate band can be stable in the center, off-center, or the
three band state is unstable.

Since the second normal stress difference is difficult to measure, we
used the Johnson-Segalman and Giesekus models to evaluate it for the
experimental examples of micellar solutions (low shear modulus) and
polymer solutions (high shear modulus). We have not used more
molecularly-motivated models; the GLAMM model is too complex to solve
under shear banding conditions
\citep{graham2003microscopic,MilnerML01}, while the Rolie-Poly model
has zero second normal stress \citep{likhtmangraham03}.

Our model may thus explain some recently-reported phenomena: for example, reducing the
rim gap delays the onset of meniscus instability \citep{sui2007iep};
a larger modulus $G$ apparently increases the
possibility of instability (when comparing wormlike micelles to entangled polymer solutions); and there are variations in how far the applied shear rate progresses along the
stress plateau \citep{BPD97}. Finally, our results allow one to rationalize the experiments on entangled polymer solutions that show a contentious mixture of shear banding and edge fracture: because of the relatively large second normal stresses of typical entangled polymer solutions, we naturally expect edge fracture to be associated with shear banding, which obviously complicates such measurements (as is apparent in the literature).

\acknowledgements
We thank Sandra Lerouge and a referee for illuminating correspondence, and the EPSRC for a DTA  studentship. 
\appendix
\section{Free Surface}\label{sec:free-surface}
Here we outline the conditions necessary for a complete calculation of the free surface problem, which would determine the base state, including secondary flows, from which a true instability could be calculated. We consider steady creeping flow for which 
$\nabla \cdot \ten{T} = 0$.
Further, for planar Couette flow uniform in the $\hat{\vec{x}}$ direction, stress gradients in $\hat{\vec{x}}$ are zero.
The free surface is subject to the boundary condition
\begin{align}
\ten{T}\cdot\hat{\vec{n}}-(-p_a\hat{\vec{n}})&=(\gamma_s \nabla\cdot\hat{\vec{n}}) 
\hat{\vec{n}}-\nabla\!_s \gamma_s\;,
\label{eqn:bc}\\
\noalign{\noindent where}
\hat{\vec{n}}&=\frac{\hat{\vec{z}}-h^\prime \hat{\vec{y}}}{\sqrt{1+h^{\prime2}}}\;,
\label{eqn:nhat}\\
\noalign{\noindent and}
\nabla\cdot\hat{\vec{n}}&=\frac{-h^{\prime\prime}}{\left(\sqrt{1+h^{\prime2}}\right)^3}
\label{eqn:gradnhat}
\end{align}
is the curvature and $h\equiv h(y)$. We take the surface tension to be uniform so $\nabla\!_s \gamma_s=0$.

\noindent Eqs. \eqref{eqn:bc}, \eqref{eqn:nhat} and \eqref{eqn:gradnhat} give
\begin{align}
\hat{\vec{x}}\; : & &\qquad -h^\prime {\rm T}_{xy}+ {\rm T}_{xz} 
\hskip-2.7truecm&&&=0\;,\\
\hat{\vec{y}}\; :  &&\qquad -h^\prime {\rm T}_{yy}+ {\rm T}_{yz}-h^\prime p_a 
\hskip-2.7truecm&&&=-h^\prime(\gamma_s \nabla\cdot\hat{\vec{n}})\;,\\
\hat{\vec{z}}\; : &&\qquad -h^\prime {\rm T}_{zy}+ {\rm T}_{zz} + p_a 
\hskip-2.7truecm&&&=\gamma_s \nabla\cdot\hat{\vec{n}}\;.
\end{align}
These three conditions determine the relation between the local curvature of the meniscus $h'(y)$ and the stress components of the fluid evaluated at the meniscus. 
As $h^\prime\rightarrow 0$ the free surface becomes flat and we recover simple shear conditions. However, generally $h^\prime$ can range from $-\infty$ to $+\infty$ as the shape of the meniscus changes, and this describes the limitation of our calculation.

The remaining conditions are no flux through the free surface,  
$\vec{v}\cdot\hat{\vec{n}}=0$; a stationary plate  
$\vec{v}(x,0,z)=\big(0,0,0\big)$; and a moving plate 
$\vec{v}(x,W,z)=\big(V,0,0\big)$. Far from the meniscus we require  
$\vec{v(z\rightarrow-\infty})=\big(v_x(y),0,0\big)$,and the flow must reduce to simple shear flow given by  uniform ${\rm T}_{yy}$ and ${\rm 
T}_{xy}$, with${\rm T}_{xz}={\rm T}_{zx}={\rm T}_{yz}={\rm T}_{zy}=0$.
From incompressibility $\nabla \cdot \vec{v} = 0$, and the total stress is given by $\ten{T} = -p \Id + 2 \eta \ten{D} + \ten{\Sigma}$, where $\ten{D} \equiv \half\left[\nabla\vec{v} + (\nabla \vec{v})^T\right]$ and $\ten{\Sigma}$ depends on the particular model.

It is a challenging problem  to find the flow and normal stresses at the free surface while also finding the shape of the free surface, which is generally not circular. For example, \citet{tanner1983shear} assumed an initial semi-cylindrical crack in their calculation of edge fracture, while \citet{Lodge64} showed that a spherical surface leads to a solution consistent with mechanical balance. Neither calculation will apply for a shear banded state where there must, by necessity, be different curvatures in the two bands due to different normal stresses. Additionally, there is the complication of two (or more) bands with different shear rates in the bulk.

\section{Constitutive equations}\label{sec:const-equat}
In this Appendix we present the homogeneous steady states of the two
constitutive models we have used to fit the data in
Section~\ref{sec:application-model}.  We fit the total shear stress
$T_{xy}=G\Sigma_{xy} + \eta\dot{\gamma}$ to the measured flow curves
of shear stress as a function of shear rate. Note that the flow curves
include the stress plateau, while the constitutive curves are
non-monotonic. We fix the position of the stress
plateau based on the experiments, rather than fitting to a non-local model \citep{lu99}. This allows us to
extract the second normal stress difference in the two shear bands, as
the values $N_2(\dot{\gamma}_1)$ and  $N_2(\dot{\gamma}_2)$   are determined by the shear rates $\do{\gamma}_1$ and $\dot{\gamma}_2$ in the coexisting shear bands, given by the intersection of the stress plateau with the two stable branches of the constitutive
curve.\\[5truept]
\begin{widetext}
\subsection{Johnson-Segalman model}
In the diffusive Johnson-Segalman (JS) model the viscoelastic stress 
$\ten{\Sigma}$ is assumed to obey \citep{johnson77,olmsted99a}
\begin{align}
(\partial_t + \vec{v}\cdot \nabla)
\ten{\Sigma}+(\ten{\Omega}\ten{\Sigma} - \ten{\Sigma}
\ten{\Omega})-a(\ten{D}\ten{\Sigma} + \ten{\Sigma}\ten{D})
+\frac{1}{\tau} \ten{\Sigma} &=  
  2 \frac{\mu}{\tau}\ten{D} +{\mathcal D} \nabla^2
  \ten{\Sigma},
  \label{eqn:jsmodel}
\end{align}
\end{widetext}
where $\tau$ is the linear relaxation time and $a$, which satisfies
$-1<a<1$, parametrizes slip of the polymer relative to the local flow
field.  The JS fluid has a Newtonian viscosity $\eta$ due to the
solvent and a polymer viscosity $\mu$, which is related to the
characteristic elastic modulus $G$ by $\mu=G\tau$. We define the viscosity ratio 
\begin{equation}
  \label{eq:viscosityratio}
   \epsilon=\frac{\eta}{\mu}.
\end{equation}
Banding occurs for $0<a<1$ and $0<\epsilon<\frac{1}{8}$.

Variables are made non-dimensional according 
\begin{align}
\hat{\ten{\Sigma}}&=\frac{\ten{\Sigma}}{G},&
\hat{\dot{\gamma}} &\equiv\tau\dot{\gamma},
\end{align}
in terms of which the total stress, Eq.~(\ref{eqn:totalstress}), is
expressed as $\hat{\ten{T}} = -\hat{p}\ten{I} + 2\epsilon\hat{\ten{D}}
+ \hat{\ten{\Sigma}}$.  We will use the same scaling for the
Giesekus model. For planar Couette flow the steady state
solution to the homogeneous JS equation is \citep{larson88,Bird}
\begin{subequations}
  \begin{align}
    \hat{\Sigma}_{xy}
    &=\frac{\hat{\dot{\gamma}}}{1+(1-a^2)\hat{\dot{\gamma}}^2},&
    \hat{\Sigma}_{zz} &=0\\
    \hat{\Sigma}_{yy} &=\frac{(-1+a)\hat{\dot{\gamma}}^2}{1 +
      (1-a^2)\hat{\dot{\gamma}}^2},
  \end{align}
\end{subequations}
in terms of which the second normal stress difference is given by $\hat{N}_2 =
\hat{\Sigma}_{yy}$. 
\subsection{Giesekus Model}
\begin{widetext}
In the diffusive Giesekus model the viscoelastic stress 
$\ten{\Sigma}$ is assumed to obey 
 \citep{Gies82,Helg}
\begin{align}
(\partial_t + \vec{v}\cdot \nabla)
\ten{\Sigma}+(\ten{\Omega}\ten{\Sigma} - \ten{\Sigma}
\ten{\Omega})-(\ten{D}\ten{\Sigma} + \ten{\Sigma}\ten{D})
+\frac{1}{\tau}\ten{\Sigma}  &=  
  2 \frac{\mu}{\tau}\ten{D} - \frac{a}{\mu}
  \ten{\Sigma}^2+{\mathcal D} \nabla^2 
  \ten{\Sigma},
  \label{eqn:giesekus}
\end{align}
Here the non-monotonic behaviour comes not from slip, but from a
non-linear relaxation term parametrized by $a$. The analytical
solutions for planar Couette flow are given by
\citep{Bird,Gies82}. 
\begin{subequations}
  \begin{align}
    \hat{N}_2 &=\frac{1-\Lambda}{1+(1-2a)\Lambda}\;,
    \text{} \quad \hat{\Sigma}_{xy}
    =\frac{(1-\hat{N}_2)^2\hat{\dot{\gamma}}}{1+(1-2a)\hat{N}_2}\;, \\[5truept]
    \noalign{\text{where}}\nonumber\\[-6truept]
    \Lambda^2
    &=\frac{1}{8a (1-a) \hat{\dot{\gamma}}^2}\left[\sqrt{1 + 16a(1-a)
        \hat{\dot{\gamma}}^2}-1\right]. 
  \end{align}
\end{subequations}
A non-monotonic constitutive relation, and hence shear banding, is
possible for $a>\frac{1}{2}$  and $\epsilon<(2a-1)^2/(2a)$.
\section{Three Band Configuration}\label{sec:three-band-conf}
\subsection{Meniscus profiles}
In this Appendix we compute the configurations for three bands, with a
high shear band sandwiched between two low shear rate bands.
Motivated by \citeauthor{BritCall97c}'s experiments, we assume a
central high shear rate band of width $w_2$ and peripheral low shear
rate bands $w_1$ and $w_3$, where $w_1+w_2+w_3=W$.  An equivalent
construction applies to a central low shear rate band.

The height of the surface $h(y)$ is given in the three regions by (Fig.~\ref{fig:TBP})
\begin{align}
h_1(y) &= R_1 \left[\sqrt{1-\left(\dfrac{y}{R_1} + \cos\phi_1\right)^2} -
\sin\phi_1\right] \\[4ex] 
h_2(y) &= R_2\sqrt{1-\left(\dfrac{y-w_1}{R_2} + \frac{w_1}{R_1} +
  \cos\phi_1\right)^2} + (R_1-R_2)\sqrt{1 - \left(\dfrac{w_1}{R_1} +
  \cos\phi_1\right)^2}-R_1\sin\phi_1 \\[4ex] 
h_3(y) &= R_1\left[\sqrt{1 - \left(\dfrac{y-W}{R_1} - \cos\phi_2\right)^2} -
\sin\phi_2\right] + H\;. 
\label{eqn:CPheight}
\end{align}
Continuity of $h$ at $y=w_1$ and $y=w_1+w_2$ requires the difference in height $H$ across the gap to be given by
\begin{equation}
H = (R_1-R_2) \left[\sqrt{1-\left(\frac{w_1}{R_1}+\cos\phi_1\right)^2}
- \sqrt{1-\left(\frac{w_1}{R_1}+ 
  \frac{w_2}{R_2}+\cos\phi_1\right)^2}\right] + R_1\left(\sin\phi_2 -
\sin\phi_1\right)\;,  
\label{eqn:CPtopC}   
\end{equation}
and continuity of $h^\prime$ at $y=w_1$ and $y=w_1+w_2$ requires
\begin{equation}
\frac{w_1}{R_1}+\frac{w_2}{R_2}+\frac{w_3}{R_3}=-\cos\phi_1-\cos\phi_2\;.
\label{eqn:CPtopA}
\end{equation}
The normal stress balance conditions (Eq.~\ref{eqn:deltaN})
require $R_3=R_1$.  The width of the high shear rate band is given by
$\hat{w}_2=(\dot{\gamma}_{\textrm{app}} - \dot{\gamma}_1)/(\dot{\gamma}_2 -
\dot{\gamma}_1)$, while 
$w_1$ and $w_3$ can, in principle, take any values as long as they
satisfy
$\hat{w}_1+\hat{w}_3=1-\hat{w}_1$. 
Eqs.~(\ref{eqn:deltaN}) and (\ref{eqn:CPtopA}) lead to
\begin{equation}
\hat{R}_1=\frac{1}{-\cos\phi_1-\cos\phi_2 - \hat{w}_2A} \quad
\text{and} \quad \hat{R}_2=\frac{1}{-\cos\phi_1-\cos\phi_2 +
  (\hat{w}_1+\hat{w}_3)A}\;. 
\label{eqn:CPtopB}
\end{equation}
\begin{figure}[tb]
\begin{center} 
\includegraphics[width = 1\textwidth]{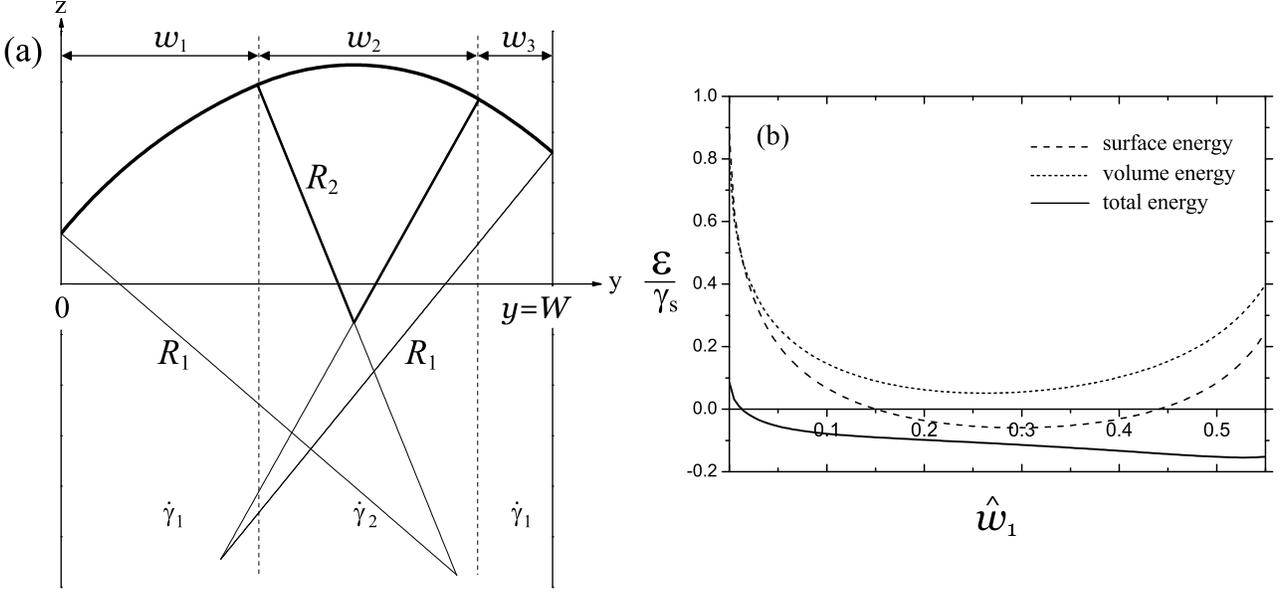}
\end{center}
\caption{(a) Profile of the meniscus for the three-band configuration for $\hat{w}_2=0.45$ and 
$\hat{w}_1=0.4$, with contact angles  $\phi_1=140^{\circ}$ and $\phi_2=130^{\circ}$, in which the outer bands have one shear rate $\dot{\gamma}_1$ and the inner band has a different shear rate.  Normal  stress balance ensures that the two outer bands have
  equal radii of curvature. The width $w_2$ of the
  inner band is given by $\hat{w}_2 =
  (\dot{\gamma}_{\textrm{app}} - \dot{\gamma}_1) / (\dot{\gamma}_2 -
  \dot{\gamma}_1)$. (b) Energy as a function of $\hat{w}_1$ showing the separate contributions from surface tension (surface) and pressure (volume). The stable configuration occurs for $\hat{w}_1\approx 0.52$.}
\label{fig:TBP}
\end{figure} 

\subsection{Meniscus Integrity}
The conditions for integrity of the surface,
 $h^\prime(w_1)<\infty $ and $h^\prime(w_1+w_2)<\infty$,
lead to  (from Eqs.~\ref{eqn:CPtopB})
\begin{subequations}
\begin{gather}
\left|\left(1 - \hat{w}_1\right)\cos\phi_1 - \hat{w}_1\cos\phi_2 -
  \hat{w}_1\hat{w}_2A\right| <1,\\ 
\intertext{}
\left|\left[1  -\left(\hat{w}_1+\hat{w}_2\right)\right]\cos\phi_1
  -(\hat{w}_1 +  \hat{w}_2)\cos\phi_2 + \hat{w}_2\left[1  -\left(\hat{w}_1 +
    \hat{w}_2\right)\right]A\right| <1. 
\end{gather}
\label{eqn:CPStabConditions}
\end{subequations}
\end{widetext}
These conditions apply at both interfaces of the central high shear
rate band, and must be satisfied simultaneously for the meniscus to maintain integrity. The contact angles $\phi_1$, $\phi_2$ and the distortion
parameter $A$ are properties of the fluid, while the width of the high
shear rate band $w_2$ is determined by the applied shear rate. 

Although a symmetric solution $w_1=w_3$ is appealing, we will consider
the more general case where the two low shear rate bands need not be
the same size. We will see that a simple stability analysis can lead
to symmetry breaking to a non-symmetric band configuration.  As has
already been mentioned, stability should strictly be determined by
dynamical considerations. However, since we are only treating the
mechanics of the meniscus, we will study the three band configuration
by constructing an energy function, which roughly corresponds to the
work needed to establish an interface, and whose gradient specifies a
generalized force that should vanish for mechanical stability.

The mechanical energy of the surface, per unit length in the flow
direction, is given by
\begin{equation}
 {\cal E}(w_1,w_2,A,\phi_1,\phi_2) = \gamma_s \int_0^{W} \sqrt{1+h'^2}
 \; \text{d}y\;-\;\Delta p\int_0^{W} h\; \text{d}y\;, 
\label{eqn:SurfaceEnergy}
\end{equation}
which comprises the surface tension and the work done in deforming
against the pressure differences across the interface. We take ${\cal
  E}=0$ at the onset of banding.  For a given width of the high shear
rate band $w_2$ we determine the position of the band by minimising
${\cal E}$ with respect to $w_1$.  We will only consider the equal
contact angle case, $\phi_1=\phi_2=\phi$. Stable configurations are
defined as those for which the energy function ${\cal E}$ is a
local minimum with respect to changing the sizes $w_1$ and $w_2$  of the two
outer shear bands at fixed $w_2$.

\subsection{Different three-band configurations}
 Three configurations of the three band model minimize
the energy, as demonstrated in Figs.~\ref{fig:Profiles}.
\begin{enumerate}
\item[(i)]
central high shear rate band ($w_1=w_3$);
\item[(ii)]
off center high shear rate band ($w_1\neq w_3$);
\item[(iii)]
collapsed low shear rate band ($w_1=0$, or $w_3=0$), corresponding to
two shear bands.
\end{enumerate}
We can show that, for equal angles, there will be an extremum at $w_1=\frac{1}{2}(1-w_2)$. The high shear rate band remains central so long as $\displaystyle{\frac{\partial ^2{\cal E}}{\partial w_1^2}\Bigg|_{\frac{1}{2}(1-w_2)}}\!\!\!\!\!\!\!\!\!\geq0$.  Equality defines  $w^{\ast}_2$, so that the condition for stability is $w_2<w^*_2$ where $w^*_2$
satisfies \vskip0.2truecm 
\begin{widetext}
\begin{equation}
\frac{1}{\sqrt{1 - \left(\frac{w_2^*}{2R_2}\right)^2}}
-\left[1 +
  \frac{A}{2\cos\phi}\frac{w_2^*}{W}\frac{R_1}{R_2}\right] +
\left(\frac{R_1}{R_2}\right)\frac{\sqrt{1 -
    \left(\frac{w_2^*}{2R_2}\right)^2}}{\sqrt{1 - 
    \left(\cos\phi\:\frac{w_2^*}{W}\right)^2}}=0,
\label{eqn:w2critical}
\end{equation}
\end{widetext}
together with the integrity conditions of equation
(\ref{eqn:CPStabConditions}). Further, for given distortion parameter
$A$ and high shear rate band width $w_2$, a three band configuration
may satisfy the integrity conditions of
Eq.~(\ref{eqn:CPStabConditions}). At a slightly greater value of $w_2$
(or applied shear rate) the three band configuration becomes unstable
and the fluid collapses into a two-band states, with integrity
governed by equation (\ref{eqn:StabConditions}).

\begin{figure*}[!htb]
\begin{center}
\includegraphics[width = 1\textwidth]{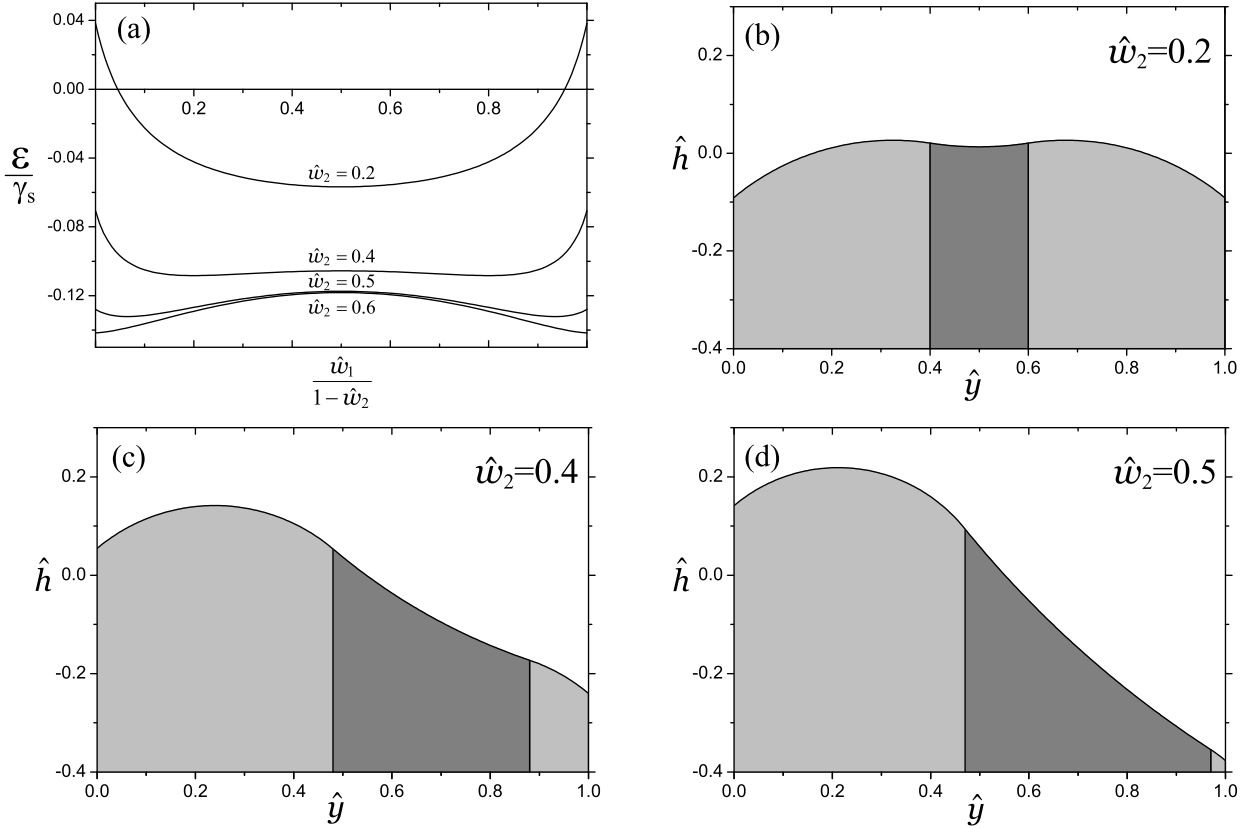}
\end{center}
\caption{Total energy curves (a) and mensci shapes (b,c,d) for distortion parameter $A=-3.5$ and
  contact angle $\phi=130^\circ$.  For widths $\hat{w}_2<0.343$ of
  the high shear rate bands (a), the central high shear rate band is
  stable.  For $\hat{w}_2>0.343$ the high shear rate band is
  no longer stable at the central position but stabilises such that
  there is a wide low shear rate band and a narrow low shear rate band
  close to the plate (c,d). For $\hat{w}_2>0.579$ the fluid collapses
  into the two band configuration.  The meniscus profiles are shown for $\hat{w}_2=0.2, 0.4, 0.5$, and the complete set of states for all contact angles is shown in Fig.~\ref{fig:threebandstates}.}
\label{fig:Profiles}
\end{figure*}

Fig.~\ref{fig:Profiles} shows the energy curves and some meniscus shapes for the distortion
parameter $A=-3.5$ and equal contact angles $\phi=130^{\circ}$, as a
function of increasing the width of the central high shear rate band
$\hat{w}_2$.  For $\hat{w}_2/(1-\hat{w}_2)=0.5$ the low shear material
splits into two equal size bands surrounding the high shear rate
band. Upon increasing $\hat{w}_2$ the symmetric three band
configuration becomes unstable to an asymmetric configuration (for
$\hat{w}_2>0.343$), and finally unstable to a two band configuration
when the energy minima lie at either $\hat{w}_1=0$ or
$\hat{w}_1=1-\hat{w}_2$. 

Fig.~\ref{fig:threebandstates} shows all loci of states whose meniscus integrity we can estimate using this method for $A=-3.5$. This figure shows the regions, in Fig.~\ref{fig:Fig6}, in which two bands fracture at different ends of the stress plateau depending on whether the contact angle is closer to wetting or non-wetting angles. In addition, there are accompanying regions in which the inner of  three bands is either symmetric or non-symmetric. In this particular example, the larger contact angles ($\phi>90^{\circ}$) correspond to a central high shear rate band, while the smaller contact angles ($\phi<90^{\circ}$) correspond to a central low shear rate band. There are four distinct classes of regions of for which the meniscus retains integrity:  (1)  two bands, (2)  three bands with a central symmetric band, (3)  either three symmetric bands or two bands, or (4) either three asymmetrically distributed bands or two bands. In the last two cases where two band configurations allow meniscus integrity we expect factors such as flow history, flow geometry (\textit{e.g.} cylindrical Couette vs cone and plate) and boundary conditions to determine the selected configuration.
\begin{figure*}[!htb]
\includegraphics[width = 1\textwidth]{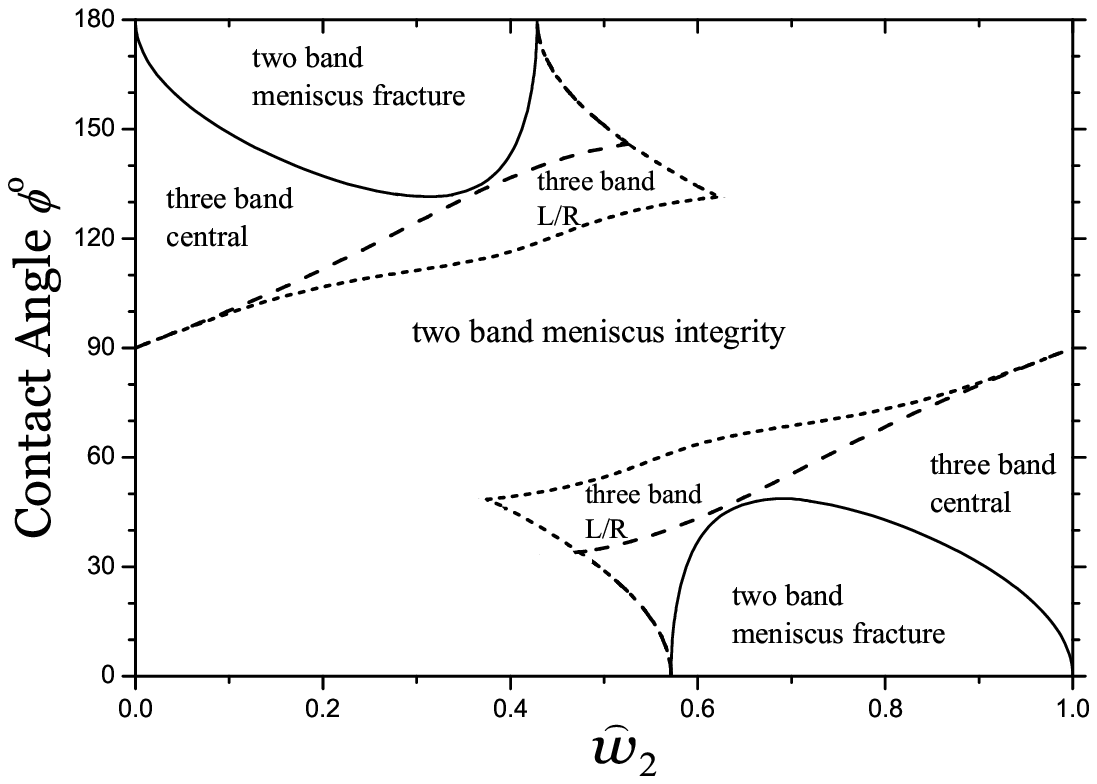}
\caption{Integrity diagram for  $A=-3.5$, show regions  in which two or three bands allow meniscus integrity as a function of contact angle and the fraction of material $\hat{w}_2$ in the middle band. For $\phi>90^{\circ}$ ($\phi<90^{\circ}$) the central band adopts the  high (low) shear rate. }
\label{fig:threebandstates}
\end{figure*}

\citet{Britton.Callaghan99} found the central high shear rate band to be
 stable and to increase in width as the applied shear rate
increased (Fig.~\ref{fig:BritCal5}). At
$\dot{\gamma}=7\;\text{s}^{-1}$ the high shear rate band is stable in
a central position consistent with Fig.~\ref{fig:Profiles}b,
while at $\dot{\gamma}=10.7\;\text{s}^{-1}$ the high shear rate band
is no longer central but in an off-center stable position consistent
with Fig.~\ref{fig:Profiles}c. This would be consistent with the state diagram in Fig.~\ref{fig:threebandstates}, for $\phi\gtrsim120^{\circ}$.
\begin{figure*}[!htb]
\begin{center}
\includegraphics[width = 1\textwidth]{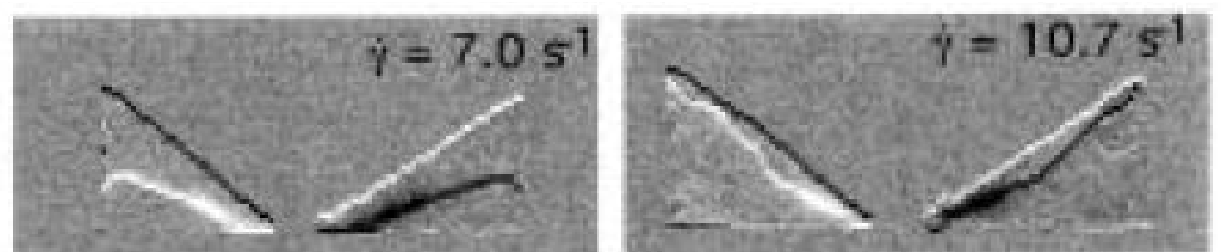}
\end{center}
\caption{NMR images of a worm-like micellar solution undergoing shear
  banding in a cone-and-plate rheometer
  (image from Fig. 10 \citet{Britton.Callaghan99}). At an applied shear rate of
  $\dot{\gamma}=7\;\text{s}^{-1}$ the high shear rate band, appearing
  as white/black, is stable and central. This would correspond to
 Figure~(\ref{fig:Profiles}b). At the higher applied
  shear rate $\dot{\gamma}=10.7\;\text{s}^{-1}$, the high shear rate
  band is stable but no longer central but positioned close to the
  moving cone. This corresponds to Figure~(\ref{fig:Profiles}c). 
 At even higher shear rates the material was expelled from the cell.}
\label{fig:BritCal5}
\end{figure*}

\subsection{Summary}
In summary, we find the following results for three bands: 
\begin{enumerate}
\item Three bands are stable in a subspace of the the parameter space, for a range of contact angles. This stability region starts with an infinitesimally small central band, and continues until the central band becomes too large, after which only a conventional two band state is possible. 
\item The stable central band is the high (low) shear rate branch when the contact angle is closer to $180^{\circ} (0^{\circ})$.
\item The central of the three bands is symmetric for smaller central bands, but is destabilized towards an off-center configuration for wider central bands, for a range of contact angles. 
\item In some regions of the phase space two-band and three-band states are simultaneously stable; in other regions the two band state is unstable, in principle towards three band states. This may explain some of the experiments of \citet{BritCall97c}, in which three band states are found.
\item We are not able to assess the relative stability of two or three band states, and we speculate that this is determined by factors such as initial conditions, boundary conditions, and the degree of curvature in the flow. For example, the highly curved flow of a wide gap Couette rheometer may have a strong preference for the two band state, as it also leads to a preference for inducing the high shear rate phase next to the inner cylinder where the shear stress is highest \citep{radulescu99a,BCAdams08}.
\end{enumerate}

%

\end{document}